\documentclass[lettersize,journal]{IEEEtran}
\usepackage{hyperref}
\usepackage{color,soul}
\usepackage{graphicx}
\usepackage[font=scriptsize]{caption}
\usepackage{subcaption}
\usepackage{booktabs}
\usepackage{siunitx}
\usepackage{enumitem}
\usepackage{amsmath}
\usepackage{amssymb}
\usepackage{multirow}
\usepackage{xurl}
\usepackage{float}
\usepackage[british]{babel}
\usepackage{csquotes}
\usepackage[mathlines]{lineno}
\usepackage{xcolor,soul}
\sethlcolor{yellow}
\usepackage{algorithm}
\usepackage{algpseudocode}
\usepackage{orcidlink}
\usepackage{makecell}
\usepackage{nicefrac}
\usepackage{tikz}
\usetikzlibrary{arrows.meta, positioning}

\usepackage{dblfloatfix}   % allow figure*/table* at page bottom and fix their ordering

\makeatletter
\renewcommand{\fnum@figure}{Fig. \arabic{figure}. }
\renewcommand{\fnum@table}{TABLE \roman{table}}
\makeatother
\begin{document}
%\linenumbers
\title{Gauge Freedom Optimization for Truncation Error Reduction in Inertial Navigation}
% authors
\author{Yaakov~Libero \orcidlink{0000-0001-5213-9985}, and Itzik Klein \orcidlink{0000-0001-7846-0654}%
\thanks{Yaakov Libero and Itzik Klein are with the Hatter Department of Marine Technologies, Charney School of Marine Sciences, University of Haifa, Haifa, Israel.}% <-this % stops a space
}
% The paper header
\maketitle
\begin{abstract}
Numerical integration plays a central role in inertial navigation systems, where sensor measurements are propagated through time to obtain orientation, velocity, and position states. The accuracy of this propagation depends on the numerical integrator type, order and step-size. Prior work showed that for second-order systems with known forcing functions, the gauge freedom in the variation of parameters technique can be exploited to reduce truncation error without modifying the integrator. However, this approach requires analytical knowledge of the forcing function, limiting its applicability in real-world systems. To address this limitation we propose the u-space methodology, a novel state mapping that generalizes the gauge freedom to systems with unknown forcing functions. The optimal gauge is derived in closed form for second-order systems and in both closed and empirical form for first-order systems. The proposed approach was evaluated through Monte Carlo simulations across four forcing functions, five sensor grades, and four Adams-Bashforth orders, as well as on a real-world inertial navigation dataset. Results show consistent error reduction across all tested conditions, with the largest gains observed in the full inertial mechanization pipeline, making the approach applicable to high-grade inertial systems, where truncation error constitutes a larger share of the error budget, and to aided low-cost systems with high-rate updates, where propagation spans only short inter-update intervals.
\end{abstract}
\IEEEpeerreviewmaketitle
\section{Introduction}
\label{sec:introduction}

Ordinary differential equations (ODEs) govern the system dynamics in a wide range of fields, with examples including control~\cite{ogata2010modern}, robotics~\cite{featherstone1983calculation}, signal processing~\cite{nakai2024ordinary}, physical simulation~\cite{verlet1967computer}, and navigation~\cite{giroux2003validation, zeinali2024imunet}. While analytical solutions exist for certain classes of ODEs, they are often unavailable in practice due to a variety of constraints. Those include systems in which the forcing function lacks an analytical expression or is measured in real-time from deployed sensors~\cite{hurwitz2024deep} and systems in which the analytical integral is too complex to resolve~\cite{butcher2016numerical}. In those cases, numerical methods are employed to obtain approximate solutions. These methods operate by evaluating the ODE at discrete time steps and constructing an approximation to the integral~\cite{iserles2009first} with single-step methods, such as the Runge-Kutta family~\cite{dormand1980family}, evaluating the forcing function using multiple points within each step, while multi-step methods, such as the Adams-Bashforth family~\cite{bashforth1883attempt}, or uses evaluations from previous steps in their approximation.

In the field of inertial navigation numerical integrators are employed to propagate the orientation, velocity, and position states using measurements obtained from the inertial measurement unit (IMU) and past states values \cite{titterton2004strapdown}. However, the approximation performed in the numerical integration introduces truncation error, whose magnitude depends on the integrator type, order, and step size~\cite{dahlquist1956convergence}. This becomes of special importance in GNSS-denied environments, where the system relays solely on the propagation accuracy to achieve a suitable navigation solution \cite{alghamdi2025autonomous, xu2024nonlinearity}. Multiple approaches have been developed to reduce this error. Higher-order methods reduce it by fitting higher-degree polynomials to the measurements. Richardson extrapolation techniques~\cite{richardson1911ix}, such as Romberg's method~\cite{romberg1955vereinfachte}, achieve higher effective order by combining results from multiple partition sizes. Adaptive step-size methods~\cite{butusov2021adaptive} reduce it locally by adjusting the step size to match the signal dynamics, and symplectic integrators~\cite{sanz1992symplectic} preserve geometric structure to limit long-term error growth. The common denominator of these approaches is that they treat the numerical method and the integrated expression as separate problems, focusing on improving accuracy by modifying the numerical technique or decreasing the step size without modifying the expression being integrated. However, increasing the integrator order requires caching additional evaluations, which increases memory and computational costs \cite{papez2013numerical}. Similarly, reducing the step size is often impractical in real-world systems where the sampling rate is fixed by the sensor hardware specifications \cite{al2023imu}. A different family of approaches reformulates the equation prior to numerical integration to achieve better numerical accuracy. Exponential integrators~\cite{hochbruck2010exponential} treat a linearization of the dynamics analytically through the matrix exponential and integrate only the residual numerically, while Magnus integrators~\cite{blanes2009magnus} reformulate first-order linear ODEs as the exponential of a Lie-algebraic series. Within this family, Gurfil and Klein~\cite{gurfil2006mitigating, gurfil2007stabilizing} explored a different reformulation strategy specific to second-order ODEs. Exploiting the gauge freedom that exists in the variation of parameters (VoP) technique it was demonstrated that the gauge function itself can be chosen to reduce truncation error by several orders of magnitude without modifying the numerical integrator order or step size. However, their proposed approach required analytical knowledge of the forcing function, limiting its applicability to simulations and systems with known signal.

To bridge this gap, in this paper we introduce the u-space formulation, a novel state-space approach that generalizes the gauge freedom approach to systems with unknown forcing functions. We derive the optimal gauge in closed form for second-order systems and extend the formulation to first-order systems. The contributions of this paper are:

\begin{enumerate}
\item Generalized the gauge freedom approach to systems with unknown forcing functions, with a closed-form optimal gauge for the position and velocity states. 
\item Expanded the gauge to cover first orders systems, and derived both closed and data-driven form for the quaternion states.
\item Identified and quantified the truncation error cascade mechanism in INS mechanization, showing it accounts for 10-30\% of the total error, one order of magnitude larger than the theory predicts.
\end{enumerate}

A comprehensive evaluation through Monte-Carlo simulations and a real-world inertial navigation dataset comprising 17 vehicle trajectories was made. In simulation, u-space consistently outperformed x-space, achieving the same accuracy with a lower-order integrator. In the real-world experiment, the full inertial mechanization pipeline showed 8-31\% position root mean squared error reduction after 5 seconds.

The remainder of this paper is organized as follows: Section~\ref{sec:problem} presents the problem formulation. Section~\ref{sec:proposed} derives our proposed u-space approach. Section~\ref{sec:results} describes the experimental setup and presents simulation and real-world results, followed by the conclusions in Section~\ref{sec:conclusions}.
\section{Problem Formulation}
\label{sec:problem}

\subsection{Solving $\ddot{x} = f(t)$ with Variation of Parameters}
\label{sec:vop}
We follow the standard variation of parameters (VoP) formulation as presented in a multitude of textbooks \cite{boyce2021elementary, tenenbaum1985ordinary, coddington1956theory}. Consider the second-order ordinary differential equation
\begin{equation}
    \label{eq:ode}
    \ddot{x} = f(t)
\end{equation}
where $f(t)$ is a known continuous forcing function. The homogeneous equation $\ddot{x} = 0$ has the solution
\begin{equation}
    \label{eq:homogeneous}
    x_h(t) = c_1 \cdot 1 + c_2 \cdot t
\end{equation}
spanned by the basis functions $1$ and $t$, with $c_1, c_2$ constant. In order to obtain the particular solution we promote the constants $c_1, c_2$ to time-varying functions $u_1(t), u_2(t)$
\begin{equation}
    \label{eq:vop_ansatz}
    x(t) = u_1(t) + u_2(t)\, t
\end{equation}
This ansatz introduces two unknown functions constrained by a single equation, leaving one degree of freedom. Differentiating yields
\begin{equation}
    \label{eq:x_dot_vop}
    \dot{x} = \dot{u}_1 + \dot{u}_2\, t + u_2
\end{equation}
The standard VoP employs the gauge freedom by eliminating the first-order u-terms derivatives by setting
\begin{equation}
    \label{eq:standard_gauge}
    \dot{u}_1 + \dot{u}_2\, t = 0
\end{equation}
Reducing (\ref{eq:x_dot_vop}) to 
\begin{equation}
    \dot{x} = u_2
\end{equation}
Differentiating once more and applying (\ref{eq:ode}) gives
\begin{equation}
    \label{eq:u2_dot_vop}
    \dot{u}_2 = f(t)
\end{equation}

which alongside (\ref{eq:standard_gauge}) form a system of equations. Expressing it in matrix form using the Wronskian matrix $\mathbf{W}_2$ results in the system
\begin{equation}
    \label{eq:wronskian_system}
    \underbrace{
    \begin{bmatrix} 1 & t \\ 0 & 1 \end{bmatrix}}_{\mathbf{W_2}}
    \begin{bmatrix} \dot{u}_1 \\ \dot{u}_2 \end{bmatrix}
    =
    \begin{bmatrix} 0 \\ f(t) \end{bmatrix}
\end{equation}
with the solution
\begin{equation}
    \label{eq:u_dots}
    \begin{bmatrix} \dot{u}_1 \\ \dot{u}_2 \end{bmatrix}
    =
    \begin{bmatrix} -t\, f(t) \\ f(t) \end{bmatrix}
\end{equation}
Finally, integrating yields the particular solution
\begin{equation}
    \label{eq:particular}
    x(t) = -\int t\, f(t)\, dt \;+\; t \int f(t)\, dt
\end{equation}

\subsection{Numerical Integration}
\label{sec:numerical_integration}
The particular solution (\ref{eq:particular}) requires evaluating the integral of $f(t)$ analytically. In cases where the forcing function is impractical to integrate analytically or is measured in real-time we employ a numerical integration scheme. First, the function is decomposed into two first-order equations
\begin{equation}
    \label{eq:v_dot}
    \dot{v} = f(t)
\end{equation}
\begin{equation}
    \label{eq:x_dot}
    \dot{x} = v(t)
\end{equation}
where $v(t)$ is the velocity. Both equations have the same differential form $\dot{y}(t) = h(t)$, whose integral form is
\begin{equation}
    \label{eq:integral_form}
    y_{n+1} = y_n + \int_{t_n}^{t_n + \Delta t} h(t)\, dt
\end{equation}
This integral can be approximated by a variety of numerical methods. In this paper, we approximate this integral using the Adams-Bashforth (AB) family of multistep methods \cite{iserles2009first}. For a uniform step size $\Delta t$, the first four orders are
\begin{align}
    \text{AB1:}\quad & y_{n+1} = y_n + \Delta t\, h_n \label{eq:ab1}\\
    \text{AB2:}\quad & y_{n+1} = y_n + \frac{\Delta t}{2}(3h_n - h_{n-1}) \label{eq:ab2}\\
    \text{AB3:}\quad & y_{n+1} = y_n + \frac{\Delta t}{12}(23h_n - 16h_{n-1} + 5h_{n-2}) \label{eq:ab3}\\
    \text{AB4:}\quad & y_{n+1} = y_n + \frac{\Delta t}{24}(55h_n - 59h_{n-1} \nonumber\\
    & \qquad\qquad\qquad + 37h_{n-2} - 9h_{n-3}) \label{eq:ab4}
\end{align}
where AB$k$ requires a cache of the $k$ most recent values of $h$.
The AB methods approximate the integral by fitting a polynomial of degree $k-1$ through the cached values of $h$. The local truncation error of AB$k$ is $\mathcal{O}(\Delta t^{k+1})$ and is proportional to $h^{(k)}(t)$. As a result, the numerical accuracy depends not only on the step size and the method order, but also on the behavior of $h^{(k)}(t)$, with rapidly varying functions leading to greater truncation error.

While the framework above is formulated for equations already in the form of (\ref{eq:ode}), second-order ODEs often include terms that depend on the state and its derivatives. Consider a general second-order ODE of the form
\begin{equation}
    \label{eq:general_ode}
    \ddot{x} + a\dot{x} + bx = f(t)
\end{equation}
where $a$, $b$ are constant coefficients. Isolating the highest-order derivative by moving all remaining terms to the right-hand side results in
\begin{equation}
    \label{eq:sv_reformulation}
    \ddot{x} = f(t) - a\dot{x} - bx
\end{equation}
At any discrete time step $t_n$, this expression can be approximated from the known values $x_n$ and $\dot{x}_n$, yielding a new forcing function $g(t)$
\begin{equation}
    \label{eq:f_eval}
    g(t) \approx g(t,x(t_n)) = f(t) - a\dot{x}_n - bx_n
\end{equation}
which reduces the problem to the desired form.

\subsection{Truncation Error Reduction via Gauge Selection}
\label{sec:gauge_truncation}
In deriving (\ref{eq:u_dots}), the gauge condition (\ref{eq:standard_gauge}) was originally selected in order to ease the process of finding the analytical solution. Gurfil and Klein~\cite{gurfil2006mitigating} showed that by relaxing this condition we can employ the gauge freedom to reduce the numerical truncation error. Consider the generalized gauge condition
\begin{equation}
    \label{eq:general_gauge}
    \dot{u}_1 + \dot{u}_2\, t = \Phi(t)
\end{equation}
where $\Phi(t)$ is an arbitrary smooth function, with $\Phi = 0$ recovering (\ref{eq:standard_gauge}). Re-deriving the variational system with (\ref{eq:general_gauge}) yields
\begin{equation}
    \label{eq:gauge_system}
    \begin{bmatrix} \dot{u}_1 \\ \dot{u}_2 \end{bmatrix}
    =
    \begin{bmatrix} \Phi - t\,(f - \dot{\Phi}) \\ f - \dot{\Phi} \end{bmatrix}
\end{equation}
The authors show that the analytical solution is unique by following Cauchy's existence and uniqueness theorem and is therefore invariant of the choice of $\Phi$. However, while the analytical solution is invariant for the propagated empirical solution this property does not hold. As demonstrated in Section~\ref{sec:numerical_integration}, the truncation error of an order-$k$ method depends on the higher-order derivatives of the state's expression. Since this expression now depends on $\Phi$ and its derivatives, different choices of $\Phi$ produce different truncation errors for the same analytical solution.

This observation can be exploited by optimizing $\Phi$ to eliminate the truncation error. For an order-$k$ method the local truncation error takes the form
\begin{equation}
    \label{eq:lte_general}
    \text{LTE}(y_n) = \gamma_k\, \Delta t^{k+1}\, h^{(k)}(t_n)
\end{equation}
where $\gamma_k$ is the method's error constant and $h^{(k)}$ denotes the $k$-th derivative of the state's expression and $y_n$ denotes a variable at time $t_n$. To minimize the truncation error, both $f(t)$ and $\Phi(t)$ were expressed as Fourier series. Since the coefficients of $f(t)$ are known it is possible to derive an optimal $\Phi$ by minimizing the squared truncation error with respect to these coefficients. The paper showed that for a range of second-order systems the optimized gauge reduced integration errors by three to seven orders of magnitude without modifying the integrator order or the step size.

\subsection{INS Mechanization}
\label{sec:ins_mechanization}
We follow~\cite{farrell2008aided} for the presentation of the INS mechanization equations. The attitude is represented by the unit quaternion $\mathbf{q}^n_b \in \mathbb{R}^4$, denoting the rotation from the body frame to the navigation frame, defined as a local North-East-Down (NED) cartesian frame with origin at the initial position.

Under the local navigation frame assumption, the Earth's rotation and transport rates are dropped, and the body angular velocity with respect to the navigation frame reduces to the gyroscope measurement
\begin{equation}
    \label{eq:omega_nb}
    \boldsymbol{\omega}^b_{nb} = \tilde{\boldsymbol{\omega}}^b_{ib}
\end{equation}
where $\tilde{\boldsymbol{\omega}}^b_{ib} \in \mathbb{R}^3$ is the gyroscope measurement in the body frame. The quaternion kinematics are governed by
\begin{equation}
    \label{eq:q_kinematics}
    \dot{\mathbf{q}}^n_b = \frac{1}{2}\,\mathbf{q}^n_b \otimes \boldsymbol{\omega}^b_{nb}
\end{equation}
where $\otimes$ denotes the Hamilton product. All vectors $\mathbf{v}\in\mathbb{R}^3$ entering the Hamilton product are embedded as pure quaternions $[0,\, \mathbf{v}]^T\in\mathbb{R}^4$.

The velocity vector expressed in the navigation frame, $\mathbf{v}^n \in \mathbb{R}^3$, evolves according to
\begin{equation}
    \label{eq:v_dot_ins}
    \dot{\mathbf{v}}^n = \mathbf{q}^n_b \otimes \tilde{\mathbf{f}}^b \otimes (\mathbf{q}^n_b)^* + \mathbf{g}^n
\end{equation}
where $\tilde{\mathbf{f}}^b \in \mathbb{R}^3$ is the accelerometer-measured specific force in the body frame, $(\mathbf{q}^n_b)^*$ is the quaternion conjugate, and $\mathbf{g}^n \in \mathbb{R}^3$ is the gravity vector in the navigation frame, assumed constant for the duration of the trajectory.

Finally, the position $\mathbf{p}^n \in \mathbb{R}^3$, expressed in the local navigation frame in meters, evolves as
\begin{equation}
    \label{eq:p_dot_ins}
    \dot{\mathbf{p}}^n = \mathbf{v}^n.
\end{equation}
\section{Proposed Approach}
\label{sec:proposed}
The goal of the proposed approach is to exploit the gauge freedom introduced in Section~\ref{sec:gauge_truncation} to reduce the truncation error of the inertial equations without requiring prior analytical knowledge of the forcing function. We start by applying the proposed approach to the second-order position kinematics, and then extending it further to include the first-order angular kinematics.
\subsection{Position Kinematics}
\label{sec:position_kinematics}
Following~\cite{groves2008}, the rate of change of the velocity vector expressed in the local navigation frame are governed by
\begin{equation}
    \label{eq:pk_v_dot}
    \dot{\mathbf{v}}^n = \mathbf{q}^n_b \otimes \tilde{\mathbf{f}}^b \otimes (\mathbf{q}^n_b)^* + \mathbf{g}^n
\end{equation}
In local coordinate frame, the position vector rate of change is
\begin{equation}
    \label{eq:pk_p_dot}
    \dot{\mathbf{p}}^n = \mathbf{v}^n
\end{equation}
Differentiating (\ref{eq:pk_p_dot}) and substituting (\ref{eq:pk_v_dot}) yields the second-order equation
\begin{equation}
    \label{eq:pk_2nd_order}
    \ddot{\mathbf{p}}^n = \mathbf{f}_p(t)
\end{equation}
where the forcing function $\mathbf{f}_p \in \mathbb{R}^3$ is
\begin{equation}
    \label{eq:pk_forcing}
    \mathbf{f}_p(t) = \mathbf{q}^n_b \otimes \tilde{\mathbf{f}}^b \otimes (\mathbf{q}^n_b)^* + \mathbf{g}^n
\end{equation}
At each step $\mathbf{f}_p(t_n)$ is evaluated from the IMU measurements and the current quaternion, allowing it to be treated as a known forcing function. Let $j \in \{N, E, D\}$ index the navigation-frame component, with $p_j$ and $f_{p,j}$ denoting the $j$-th components of $\mathbf{p}^n$ and $\mathbf{f}_p$, respectively. The scalar form of (\ref{eq:pk_2nd_order}) is
\begin{equation}
    \label{eq:pk_scalar_ode}
    \ddot{p}_j = f_{p,j}(t)
\end{equation}
with the homogeneous form $\ddot{p}_j = 0$. \newline Its solution is given by
\begin{equation}
    \label{eq:pk_homogeneous}
    p_{j,h}(t) = c_1 \cdot 1 + c_2 \cdot t
\end{equation}
where $c_1, c_2 \in \mathbb{R}$ are constants. To obtain the particular solution we promote $c_1, c_2$ to time-varying functions $u_{p,1,j}(t)$ and $u_{p,2,j}(t)$ as follows
\begin{equation}
    \label{eq:pk_ansatz}
    p_j(t) = u_{p,1,j}(t) + u_{p,2,j}(t)\, t
\end{equation}
This introduces two unknown functions constrained by a single equation, leaving one degree of freedom. Differentiating (\ref{eq:pk_ansatz}) once yields
\begin{equation}
    \label{eq:pk_x_dot}
    \dot{p}_j = \dot{u}_{p,1,j} + \dot{u}_{p,2,j}\, t + u_{p,2,j}
\end{equation}
The structure of (\ref{eq:pk_x_dot}) allows us to employ the gauge freedom by imposing a generalized gauge condition
\begin{equation}
    \label{eq:pk_gauge}
    \Phi_{p,j}(t) = \dot{u}_{p,1,j} + \dot{u}_{p,2,j}\, t
\end{equation}
where $\Phi_{p,j}(t)$ is an arbitrary smooth function.
\newline Substituting (\ref{eq:pk_gauge}) into (\ref{eq:pk_x_dot}) gives
\begin{equation}
    \dot{p}_j = \Phi_{p,j} + u_{p,2,j}
\end{equation}
Differentiating once more, isolating $\dot{u}_{p,2,j}$, and applying (\ref{eq:pk_scalar_ode}) results in
\begin{equation}
    \dot{u}_{p,2,j} = f_{p,j} - \dot{\Phi}_{p,j}
\end{equation}
Which together with (\ref{eq:pk_gauge}) forms the following system:
\begin{equation}
    \label{eq:pk_vop_system}
    \begin{bmatrix} \dot{u}_{p,1,j} \\ \dot{u}_{p,2,j} \end{bmatrix}
    =
    \begin{bmatrix} \Phi_{p,j} - t\,(f_{p,j} - \dot{\Phi}_{p,j}) \\ f_{p,j} - \dot{\Phi}_{p,j} \end{bmatrix}
\end{equation}
Our goal is to choose $\Phi_{p,j}$ such that the reconstructed truncation error is minimized. Since $p_j$ is reconstructed from $u_{p,1,j}$ and $u_{p,2,j}$ via (\ref{eq:pk_ansatz}), its local truncation error is
\begin{equation}
    \label{eq:pk_lte_recon}
    \text{LTE}(p_j)_n = \text{LTE}(u_{p,1,j})_n + \text{LTE}(u_{p,2,j})_n\cdot t_{n+1}
\end{equation}
where $t_n$ is the time at discrete step $n$ with step-size $\Delta t$. Computing the LTE of the u-space states requires their $(k+1)$ derivatives, which are given by
\begin{equation}
    \label{eq:pk_u1_deriv}
    u_{p,1,j}^{(k+1)} = \Phi_{p,j}^{(k)} - t\bigl(f_{p,j}^{(k)} - \Phi_{p,j}^{(k+1)}\bigr) - k\bigl(f_{p,j}^{(k-1)} - \Phi_{p,j}^{(k)}\bigr)
\end{equation}
\begin{equation}
    \label{eq:pk_u2_deriv}
    u_{p,2,j}^{(k+1)} = f_{p,j}^{(k)} - \Phi_{p,j}^{(k+1)}
\end{equation}
Substituting (\ref{eq:pk_u1_deriv})-(\ref{eq:pk_u2_deriv}) into (\ref{eq:pk_lte_recon}) and collecting terms yields
\begin{equation}
    \label{eq:pk_lte_local}
    \text{LTE}(p_j)_n = \gamma_k\,\Delta t^{k+1}\bigl[(1+k)\,\Phi_{p,j}^{(k)} - k\,f_{p,j}^{(k-1)}\bigr]
\end{equation}
Setting $\text{LTE}(p_j)_n$ to zero gives the optimality condition
\begin{equation}
    \label{eq:pk_phi_constraint}
    \Phi_{p,j}^{(k)}(t) = \frac{k}{k+1}\, f_{p,j}^{(k-1)}(t)
\end{equation}
Integrating both sides $(k-1)$ times yields
\begin{equation}
    \label{eq:pk_phi_constraint_integral}
    \dot{\Phi}_{p,j} = \tfrac{k}{k+1}\, f_{p,j}(t) + P_{(k-2)}(t)
\end{equation}
where $P_{(k-2)}(t)$ is a polynomial formed from the integration constants. Since $\Phi_{p,j}$ is arbitrarily  chosen, we set $P_{(k-2)}(t)$ to zero, reducing the constraint (\ref{eq:pk_phi_constraint_integral}) to 
\begin{equation}
    \label{eq:pk_phi_dot}
    \dot{\Phi}_{p,j} = \frac{k}{k+1}\, f_{p,j}(t)
\end{equation}
Substituting (\ref{eq:pk_phi_dot}) into (\ref{eq:pk_vop_system}) and including $\Phi_{p,j}$ as an augmented state results in the final system form:
\begin{equation}
    \label{eq:pk_uspace}
    \begin{bmatrix} \dot{u}_{p,1,j} \\ \dot{u}_{p,2,j} \\ \dot{\Phi}_{p,j} \end{bmatrix}
    =
    \begin{bmatrix} \Phi_{p,j} - \tfrac{1}{k+1}\,f_{p,j}\,t \\ \tfrac{1}{k+1}\,f_{p,j} \\ \tfrac{k}{k+1}\,f_{p,j} \end{bmatrix}
\end{equation}
The new state-space is denoted as u-space and is reconstructed back to x-space via 
\begin{equation}
    \label{eq:pk_p_recon}
    \mathbf{p}^n(t) = \mathbf{u}_{p,1}(t) + \mathbf{u}_{p,2}(t)\cdot t
\end{equation}
\begin{equation}
    \label{eq:pk_v_recon}
    \mathbf{v}^n(t) = \mathbf{u}_{p,2}(t) + \boldsymbol{\Phi}_p(t)
\end{equation}
and initialized by 
\begin{equation}
    \label{eq:pk_init}
    \begin{aligned}
        \mathbf{u}_{p,1}(t_0) &= \mathbf{p}^n(t_0) \\
        \mathbf{u}_{p,2}(t_0) &= \mathbf{v}^n(t_0) \\
        \boldsymbol{\Phi}_p(t_0) &= \mathbf{0}
    \end{aligned}
\end{equation}
where $\mathbf{u}_{p,1},\mathbf{u}_{p,2}, \boldsymbol{\Phi}_p \in \mathbb{R}^3$ are the per-channel states stacked over $j$. \newline Thus, instead of solving (\ref{eq:pk_v_dot}) and (\ref{eq:pk_p_dot}) to obtain the position and velocity states, we offer to solve (\ref{eq:pk_uspace}) and then reconstruct the position and velocity states using (\ref{eq:pk_p_recon}) and (\ref{eq:pk_v_recon}).
\subsection{Angular Kinematics}
\label{sec:angular_kinematics}
The angular velocity vector expressed in the body frame is
\begin{equation}
    \label{eq:ak_omega_full}
    \boldsymbol{\omega}^b_{nb} = \tilde{\boldsymbol{\omega}}^b_{ib} - (\mathbf{q}^n_b)^* \otimes (\boldsymbol{\omega}^n_{ie} + \boldsymbol{\omega}^n_{en}) \otimes \mathbf{q}^n_b
\end{equation}
The local navigation frame is defined as a non-rotating NED frame with origin at the initial position. Under this definition the rotation and transport rates are dropped, reducing the body angular velocity to
\begin{equation}
    \label{eq:ak_omega}
    \boldsymbol{\omega}^b_{nb} = \tilde{\boldsymbol{\omega}}^b_{ib}
\end{equation}
In turn, the quaternion kinematics are governed by
\begin{equation}
    \label{eq:ak_q_dot}
    \dot{\mathbf{q}}^n_b = \frac{1}{2}\,\mathbf{q}^n_b \otimes \boldsymbol{\omega}^b_{nb}
\end{equation}
Unlike in (\ref{eq:pk_2nd_order}), the quaternion equation is state-dependent. To treat it as a known forcing function we evaluate it at each step from the IMU measurements and the current quaternion by
\begin{equation}
    \label{eq:ak_forcing}
    \mathbf{f}_q(t, \mathbf{q}) \approx \mathbf{f}_q(t, \mathbf{q}(t_n)) \triangleq \mathbf{f}_q(t)
\end{equation}
where $t_n$ is the time at the start of the current step and $\mathbf{f}_q \in \mathbb{R}^4$ is given by
\begin{equation}
    \label{eq:ak_forcing_def}
    \mathbf{f}_q(t) = \frac{1}{2}\,\mathbf{q}(t_n) \otimes \tilde{\boldsymbol{\omega}}^b_{ib}(t)
\end{equation}
Let $i \in \{w, x, y, z\}$ index the quaternion component, with $q_i$ and $f_{q,i}$ denoting the $i$-th components of $\mathbf{q}^n_b$ and $\mathbf{f}_q$, respectively. \newline The scalar form of (\ref{eq:ak_forcing_def}) is
\begin{equation}
    \label{eq:ak_scalar_ode}
    \dot{q}_i = f_{q,i}(t)
\end{equation}
With a solution to the homogeneous form given by
\begin{equation}
    \label{eq:ak_homogeneous}
    q_{i,h}(t) = c_1
\end{equation}
where $c_1 \in \mathbb{R}$ is constant. \newline Since the homogeneous basis is one-dimensional, promoting $c_1$ to a time-varying function won't introduce a gauge freedom like in the second-order case. Therefore, to extend the same logic to the first-order case we need to introduce a second function $u_{q,2,i}(t)$ and basis $t$, resulting in
\begin{equation}
    \label{eq:ak_ansatz}
    q_i(t) = u_{q,1,i}(t) + u_{q,2,i}(t)\, t
\end{equation}
Differentiating once and applying (\ref{eq:ak_scalar_ode}) yields
\begin{equation}
    \label{eq:ak_x_dot}
    \dot{u}_{q,1,i} + \dot{u}_{q,2,i}\, t + u_{q,2,i} = f_{q,i}
\end{equation}
The gauge freedom is employed by imposing
\begin{equation}
    \label{eq:ak_gauge}
    \Phi_{q,i}(t) = \dot{u}_{q,2,i}
\end{equation}
where $\Phi_{q,i}(t)$ is an arbitrary smooth function. \newline Substituting (\ref{eq:ak_gauge}) into (\ref{eq:ak_x_dot}) and isolating $\dot{u}_{q,1,i}$ results in
\begin{equation}
    \label{eq:uq1i}
    \dot{u}_{q,1,i} = f_{q,i} - u_{q,2,i} - \Phi_{q,i}\, t
\end{equation}
Combining (\ref{eq:uq1i}) with (\ref{eq:ak_gauge}) forms the following system:
\begin{equation}
    \label{eq:ak_vop_system}
    \begin{bmatrix} \dot{u}_{q,1,i} \\ \dot{u}_{q,2,i} \end{bmatrix}
    =
    \begin{bmatrix} f_{q,i} - u_{q,2,i} - \Phi_{q,i}\, t \\ \Phi_{q,i} \end{bmatrix}
\end{equation}
Following the same procedure as in Section \ref{sec:position_kinematics}, our goal is to choose $\Phi_{q,i}$ such that the reconstructed truncation error is minimized. Since $q_i$ is reconstructed from $u_{q,1,i}$ and $u_{q,2,i}$ via (\ref{eq:ak_ansatz}), its local truncation error is
\begin{equation}
    \label{eq:ak_lte_recon}
    \text{LTE}(q_i)_n = \text{LTE}(u_{q,1,i})_n + \text{LTE}(u_{q,2,i})_n\cdot t_{n+1}
\end{equation}
Computing the LTE of the u-space states requires their $(k+1)$ derivatives, which are given by
\begin{equation}
    \label{eq:ak_u1_deriv}
    u_{q,1,i}^{(k+1)} = f_{q,i}^{(k)} - (k+1)\,\Phi_{q,i}^{(k-1)} - \Phi_{q,i}^{(k)}\, t
\end{equation}
\begin{equation}
    \label{eq:ak_u2_deriv}
    u_{q,2,i}^{(k+1)} = \Phi_{q,i}^{(k)}
\end{equation}
Substituting (\ref{eq:ak_u1_deriv})-(\ref{eq:ak_u2_deriv}) into (\ref{eq:ak_lte_recon}) and collecting terms yields
\begin{equation}
    \label{eq:ak_lte_local}
    \text{LTE}(q_i)_n = \gamma_k\,\Delta t^{k+1}\bigl[f_{q,i}^{(k)} - (k+1)\,\Phi_{q,i}^{(k-1)}\bigr]
\end{equation}
Setting $\text{LTE}(q_i)_n$ to zero gives the optimality condition
\begin{equation}
    \label{eq:ak_phi_constraint}
    \Phi_{q,i}^{(k-1)}(t) = \frac{1}{k+1}\, f_{q,i}^{(k)}(t)
\end{equation}
Integrating both sides $(k-1)$ times yields
\begin{equation}
    \Phi_{q,i} = \tfrac{1}{k+1}\, \dot{f}_{q,i}(t) + P_{(k-2)}(t)
\end{equation}
Like in the second-order case, we set the polynomial to zero, reducing the constraint to
\begin{equation}
    \label{eq:ak_phi}
    \Phi_{q,i} = \frac{1}{k+1}\, \dot{f}_{q,i}(t)
\end{equation}
However, unlike the position case where the closed form depends on $f_{p,j}$ directly, the angular closed form depends on $\dot{f}_{q,i}$, which is not directly available from the measurements. Instead, it is estimated via the finite differences method
\begin{equation}
    \label{eq:ak_fdot_estimator}
    \hat{\dot{f}}_{q,i}(t_n) = \frac{f_{q,i}(t_n) - f_{q,i}(t_{n-1})}{\Delta t}
\end{equation}
Substituting (\ref{eq:ak_phi})-(\ref{eq:ak_fdot_estimator}) into (\ref{eq:ak_vop_system}) results in the final u-space system
\begin{equation}
    \label{eq:ak_uspace}
    \begin{bmatrix} \dot{u}_{q,1,i} \\ \dot{u}_{q,2,i} \end{bmatrix}
    =
    \begin{bmatrix} f_{q,i} - u_{q,2,i} - \tfrac{1}{k+1}\,\hat{\dot{f}}_{q,i}\, t \\ \tfrac{1}{k+1}\,\hat{\dot{f}}_{q,i} \end{bmatrix}
\end{equation}
The reconstruction back to x-space is done via
\begin{equation}
    \label{eq:ak_q_recon}
    \mathbf{q}^n_b(t) = \mathbf{u}_{q,1}(t) + \mathbf{u}_{q,2}(t)\cdot t
\end{equation}
and initialized by
\begin{equation}
    \label{eq:ak_init}
    \begin{aligned}
        \mathbf{u}_{q,1}(t_0) &= \mathbf{q}^n_b(t_0) \\
        \mathbf{u}_{q,2}(t_0) &= \mathbf{0}
    \end{aligned}
\end{equation}
where $\mathbf{u}_{q,1}, \mathbf{u}_{q,2} \in \mathbb{R}^4$ are the per-channel states stacked over $i$. Thus, instead of solving (\ref{eq:ak_q_dot}) to obtain the quaternion states, we offer to solve (\ref{eq:ak_uspace}) and then reconstruct the states using (\ref{eq:ak_q_recon}).
\subsection{Algorithmic Summary}
\label{sec:u_ins_alg}
Algorithm~\ref{alg:u_ins} summarizes the inertial mechanization spanned by the u-space. At each step, the quaternion states are updated first and reconstructed to provide the rotation required by the velocity-position states.
\begin{algorithm}
\caption{Inertial Mechanization in u-space}
\label{alg:u_ins}
\begin{algorithmic}[1]
    \Statex \textbf{Input:}
    \Statex \quad $\bullet$ AB order $k$
    \Statex \quad $\bullet$ Step size $\Delta t$, step index $n$, elapsed time $\tau_n$
    \Statex \quad $\bullet$ Cached $\mathbf{f}_q, \mathbf{f}_p$ over the previous ($k-1$) steps
    \Statex \quad $\bullet$ u-space states $\{\mathbf{u}_{q,1}, \mathbf{u}_{q,2}, \mathbf{u}_{p,1}, \mathbf{u}_{p,2}, \boldsymbol{\Phi}_p\}$
    \Statex \quad $\bullet$ x-space states $\{\mathbf{q}^n_b(t_n), \mathbf{v}^n(t_n), \mathbf{p}^n(t_n)\}$
    \Statex \quad $\bullet$ IMU reading $\{\tilde{\boldsymbol{\omega}}^b_{ib}(t_n), \tilde{\mathbf{f}}^b(t_n)\}$
    \Statex \textbf{Quaternion states:}
    \State Evaluate $\mathbf{f}_q(t_n)$ using (\ref{eq:ak_forcing_def})
    \State Compute $\hat{\dot{f}}_{q,i}(t_n)$ via (\ref{eq:ak_fdot_estimator}) for each $i$
    \State Advance $\mathbf{u}_{q,1}, \mathbf{u}_{q,2}$ with AB$_k$ on (\ref{eq:ak_uspace})
    \State Reconstruct $\mathbf{q}^n_b(t_{n+1})$ via (\ref{eq:ak_q_recon}) and normalize
    \Statex \textbf{Velocity-position states:}
    \State Evaluate $\mathbf{f}_p(t_n)$ using (\ref{eq:pk_forcing})
    \State Advance $\mathbf{u}_{p,1}, \mathbf{u}_{p,2}, \boldsymbol{\Phi}_p$ with AB$_k$ on (\ref{eq:pk_uspace})
    \State Reconstruct $\mathbf{p}^n(t_{n+1}), \mathbf{v}^n(t_{n+1})$ via (\ref{eq:pk_p_recon})-(\ref{eq:pk_v_recon})
    \Statex \textbf{Finalize:}
    \State Cache $\mathbf{f}_q(t_n), \mathbf{f}_p(t_n)$ for the next AB step
    \State $\tau_{n+1} \leftarrow \tau_n + \Delta t$
    \State \Return $\mathbf{q}^n_b(t_{n+1}), \mathbf{v}^n(t_{n+1}), \mathbf{p}^n(t_{n+1})$
\end{algorithmic}
\end{algorithm}
\section{Analysis and Results}
\label{sec:results}
\begin{table*}[!tb]
    \centering
    \caption{Attitude simulation results (RMSE) [$^\circ$] for the strategic grade sensor, improvement [\%] relative to AB1.}
    \label{tab:sim_1st_results}
    \begin{tabular}{l*{5}{c}c}
        \toprule
        & \multicolumn{5}{c}{\textbf{Prediction window} $M$ \textbf{[steps]}} & \\
        \cmidrule(lr){2-6}
        & 1 & 5 & 20 & 100 & 500 & $\mathbf{M_{\text{eff}}}$ \\
        \midrule
        AB1 $x$-space & 6.75e-5 & 2.00e-4 & 7.42e-4 & 3.64e-3 & 3.63e-3 & $>$1500 ($>$15 [s]) \\
        AB2 $x$-space & 1.49e-5 & 2.13e-5 & 3.81e-5 & 9.03e-5 & 1.56e-4 & 20 (0.2 [s]) \\
        AB3 $x$-space & 1.34e-5 & 1.95e-5 & 3.56e-5 & 9.01e-5 & 1.56e-4 & 20 (0.2 [s]) \\
        AB4 $x$-space & 1.24e-5 & 1.91e-5 & 3.32e-5 & 8.99e-5 & 1.56e-4 & 20 (0.2 [s]) \\
        \midrule
        AB1 $u$-space & \textbf{1.07e-5 (84\%)} & \textbf{1.78e-5 (91\%)} & \textbf{3.15e-5 (97\%)} & \textbf{8.98e-5 (97\%)} & \textbf{1.53e-4 (96\%)} & \\
        \bottomrule
    \end{tabular}
\end{table*}
\begin{table*}[!tb]
    \centering
    \caption{Position simulation results (RMSE) [$m$] and relative improvement [$\%$] for the strategic grade sensor.}
    \label{tab:sim_2nd_results}
    \begin{tabular}{ll*{6}{c}c}
        \toprule
        & & \multicolumn{6}{c}{\textbf{Prediction window} $M$ \textbf{[steps]}} & \\
        \cmidrule(lr){3-8}
        & & 1 & 5 & 10 & 20 & 100 & 500 & $\mathbf{M_{\text{eff}}}$ \\
        \midrule
        \multirow{2}{*}{\textbf{AB1}}
        & RMSE$_x$ & 8.3e-5 & 2.4e-4 & 4.3e-4 & 7.4e-4 & 4.4e-3 & 2.2e-2 & \\
        & RMSE$_u$ & \textbf{5.8e-7 (99\%)} & \textbf{5.7e-6 (98\%)} & \textbf{2.0e-5 (95\%)} & \textbf{7.5e-5 (90\%)} & \textbf{1.9e-3 (57\%)} & \textbf{1.0e-2 (55\%)} & $>$1500 ($>$15 [s]) \\
        \midrule
        \multirow{2}{*}{\textbf{AB2}}
        & RMSE$_x$ & 6.0e-7 & 2.1e-6 & 4.1e-6 & 9.0e-6 & 5.1e-5 & 4.8e-4 & \\
        & RMSE$_u$ & \textbf{2.7e-7 (55\%)} & \textbf{7.4e-7 (65\%)} & \textbf{1.7e-6 (59\%)} & \textbf{4.0e-6 (55\%)} & \textbf{4.8e-5 (5\%)} & 4.6e-4 (4\%) & 100 (1 [s]) \\
        \midrule
        \multirow{2}{*}{\textbf{AB3}}
        & RMSE$_x$ & 3.4e-7 & 7.9e-7 & 1.7e-6 & 4.2e-6 & 5.0e-5 & 4.3e-4 & \\
        & RMSE$_u$ & \textbf{2.0e-7 (41\%)} & \textbf{7.0e-7 (11\%)} & \textbf{1.6e-6 (6\%)} & \textbf{4.0e-6 (5\%)} & 4.9e-5 (2\%) & 4.3e-4 (0\%) & 20 (0.2 [s]) \\
        \midrule
        \multirow{2}{*}{\textbf{AB4}}
        & RMSE$_x$ & 2.4e-7 & 7.2e-7 & 1.7e-6 & 4.2e-6 & 4.9e-5 & 4.3e-4 & \\
        & RMSE$_u$ & \textbf{1.5e-7 (38\%)} & \textbf{6.8e-7 (6\%)} & \textbf{1.6e-6 (6\%)} & \textbf{4.0e-6 (5\%)} & 4.8e-5 (2\%) & 4.3e-4 (0\%) & 20 (0.2 [s]) \\
        \bottomrule
    \end{tabular}
\end{table*}
\begin{table}[!b]
    \centering
    \caption{Sensor noise grades used in Monte Carlo simulations.}
    \label{tab:sensor_grades}
    \begin{tabular}{lc}
        \toprule
        Grade & $\sigma$ [m/s$^2$] \\
        \midrule
        Strategic  & 0.000694 \\
        Navigation & 0.00243 \\
        Tactical   & 0.0052 \\
        Industrial & 0.0104 \\
        Consumer   & 0.0416 \\
        \bottomrule
    \end{tabular}
\end{table}
\subsection{Datasets and Preprocessing}
\label{sec:experimental_setup}
The proposed approach was evaluated initially through Monte Carlo simulations on second and first order systems, followed by a real-world experiment on an inertial dataset.

\subsubsection{\textbf{Simulation Dataset}}
\label{sec:sim_setup}
Four forcing functions families were used to simulate accelerometer and gyroscope measurements: sinusoidal, polynomial, exponential, and damped oscillatory. At each time step, zero-mean Gaussian noise was added to the measurement to simulate five sensor grades as listed in Table~\ref{tab:sensor_grades}. For each combination of inertial signal, noise grade, and numerical method order (AB1-4), 100 Monte Carlo trials were performed for a total of 8000 runs per system order, 16000 in total. For each run, results at prediction windows of $M \in \{1, 5, 10, 20, 100, 500, 1000, 1500\}$ steps were evaluated, corresponding to durations ranging from 10 milliseconds to 15 seconds at a sampling frequency of 100~Hz.
\subsubsection{\textbf{Real-World Dataset}}
\label{sec:real_setup}
The real-world evaluation utilized the car subset of the MAGF-ID dataset~\cite{yampolsky2024multiple}, comprising 17 trajectories recorded at the University of Haifa campus. The dataset's C1 sensor configuration consists of nine Xsens DOT \cite{XDOT} IMUs arranged in a box-shaped pattern. The nine DOT measurements were averaged to form a single IMU. Ground truth (GT) was provided by an Inertial Labs MRU-P \cite{InertialLabs} unit equipped with real time kinematics (RTK) GNSS receiver.

Several preprocessing steps were applied to the raw data. These include the temporal alignment of the DOT and MRU-P recordings, debiasing the accelerometer and gyroscope measurements and high-passing the signal to reduce the noise level.

To maintain the same Monte Carlo methodology, start points were generated by sampling each trajectory 100 times, uniformly spaced across the 17 recordings for a total of 1700 trials. At each starting point, the integration was initialized from GT states and propagated forward over the prediction windows $M \in \{1, 6, 12, 24, 120, 600, 1200, 1800\}$, corresponding to durations ranging from 8.3 milliseconds to 15 seconds at a sampling frequency of 120~Hz.

The real-world experiment was structured in three phases. In Phase~1, the first-order u-space formulation (\ref{eq:ak_uspace}) was applied to the quaternion kinematics as described in Section \ref{sec:angular_kinematics}. In Phase~2, position was propagated using the second-order u-space formulation (\ref{eq:pk_uspace}) applied to the position mechanization as described in Section \ref{sec:position_kinematics}, with the body-to-navigation rotation performed using GT quaternions to isolate the direct benefits of the proposed approach. In Phase~3, the GT quaternions were replaced by the propagated quaternions from Phase~1, allowing us to examine the cumulative benefits in a fully self-contained inertial propagation pipeline.
\subsection{Performance Metrics}
\label{sec:metrics}
The primary metric used throughout this work is the root mean squared error (RMSE), computed and averaged across all trials at each prediction window $M$
\begin{equation}
    \label{eq:rmse}
    \text{RMSE}(M) = \sqrt{\frac{1}{N}\sum_{i=1}^{N}\left\| \hat{x}_i(M) - x_i(M) \right\|^2}
\end{equation}
where $N$ is the number of trials, $\hat{x}_i(M)$ is the predicted state, and $x_i(M)$ is the GT at window $M$. The RMSE was evaluated both on a per-axis basis and on an aggregate 3D quantity (3D distance for position, geodesic angle for attitude and heading). The relative improvement of the u-space formulation over standard integration (x-space) is defined as
\begin{equation}
    \label{eq:improvement}
    \text{Improvement [\%]} = \left(1 - \frac{\text{RMSE}_u}{\text{RMSE}_x}\right) \times 100
\end{equation}
Positive values indicate that u-space outperforms standard integration. Since any propagated solution will diverge eventually, a metric to quantify the practical duration of the gauge was formulated. We define the \emph{effective range} $M_\text{eff}$ as
\begin{equation}
    \label{eq:effective_range}
    M_{\text{eff}} = \max\left\{ M \;\middle|\; \text{Improvement}(M) \geq 5\% \right\}
\end{equation}
For each prediction window $M_i$ we compute the relative improvement using (\ref{eq:improvement}) and if it surpasses $5\%$ we update $M_\text{eff}=M_i$ consecutively until the largest $M_i$ where the improvement is still above $5\%$ is found.
\begin{figure}[!t]
    \centering
    \begin{subfigure}[b]{\columnwidth}
        \includegraphics[width=\textwidth]{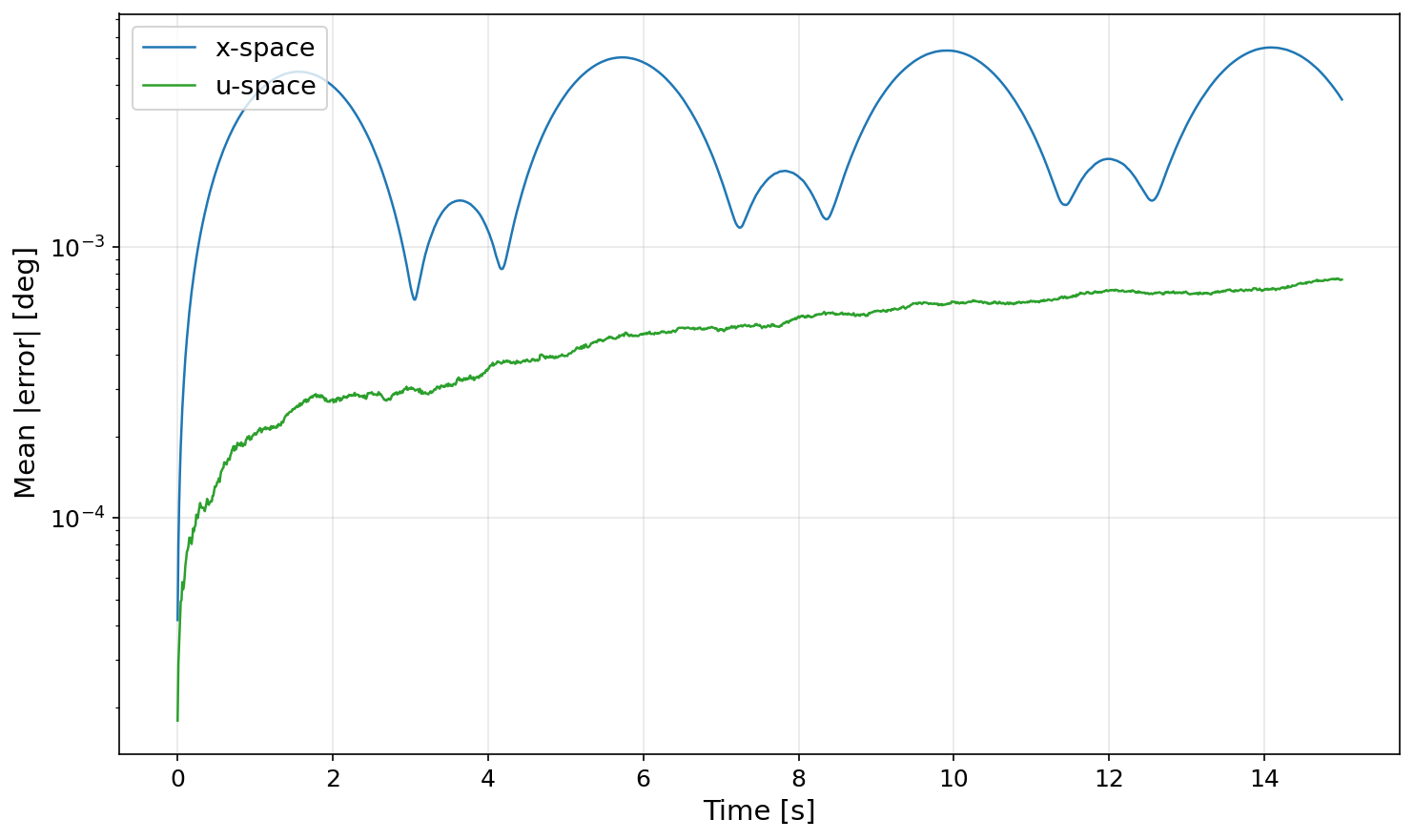}
        \caption{AB1}
        \label{fig:1st_growth_navigation_ab1}
    \end{subfigure}
    \\[0.5em]
    \begin{subfigure}[b]{\columnwidth}
        \includegraphics[width=\textwidth]{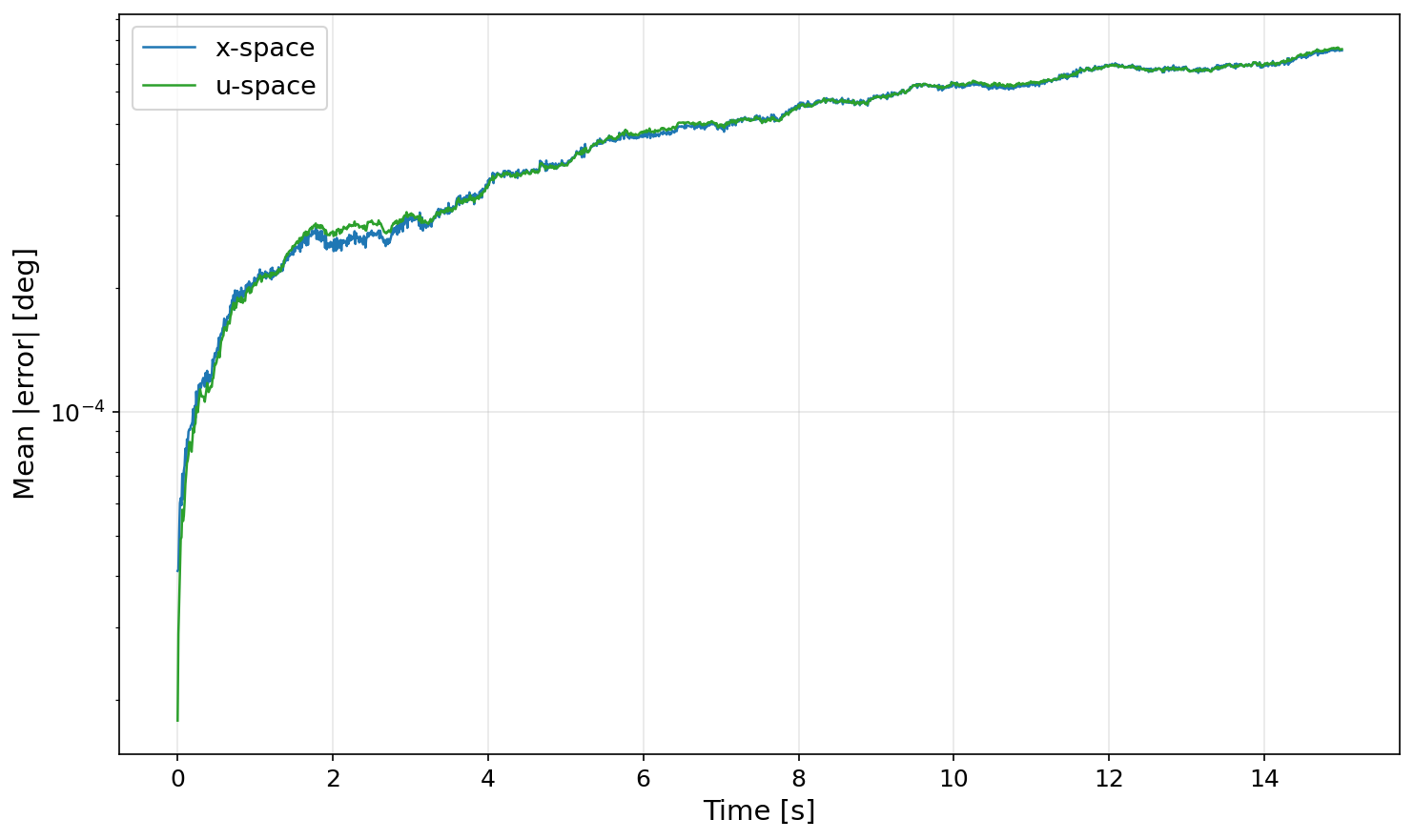}
        \caption{AB4}
        \label{fig:1st_growth_navigation_ab4}
    \end{subfigure}
    \caption{First-order simulation error growth for the navigation grade sensor for AB1 (a) and AB4 (b) x-space methods, compared to u-space AB1. AB1(x) maintains a persistent offset due to its large truncation error, while AB1(u) is able to maintain its accuracy even when compared to AB4(x).}
    \label{fig:1st_growth_consumer}
\end{figure}
\begin{figure}[!t]
    \centering
    \begin{subfigure}[b]{\columnwidth}
        \includegraphics[width=\textwidth]{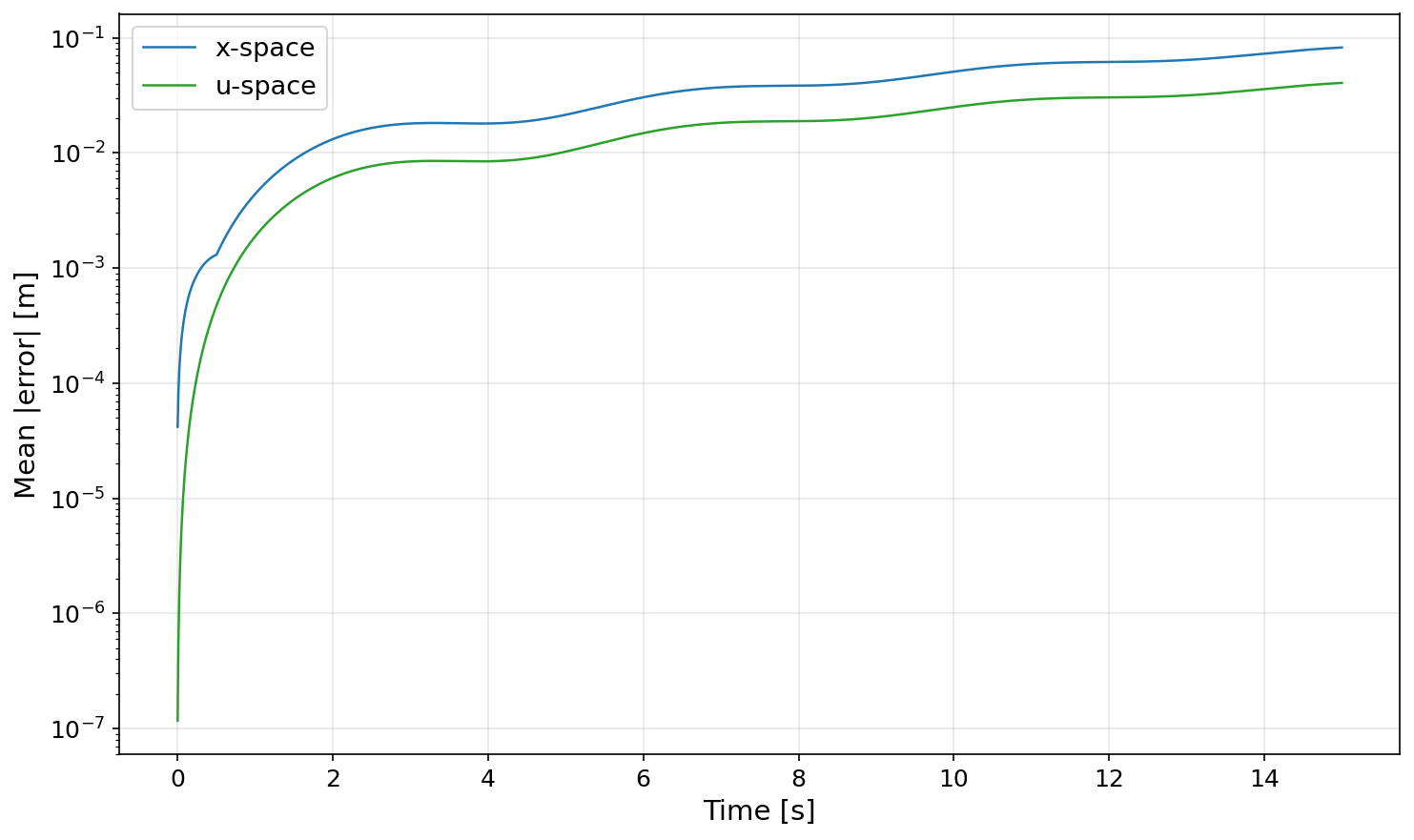}
        \caption{AB1}
        \label{fig:2nd_growth_consumer_ab1}
    \end{subfigure}
    \\[0.5em]
    \begin{subfigure}[b]{\columnwidth}
        \includegraphics[width=\textwidth]{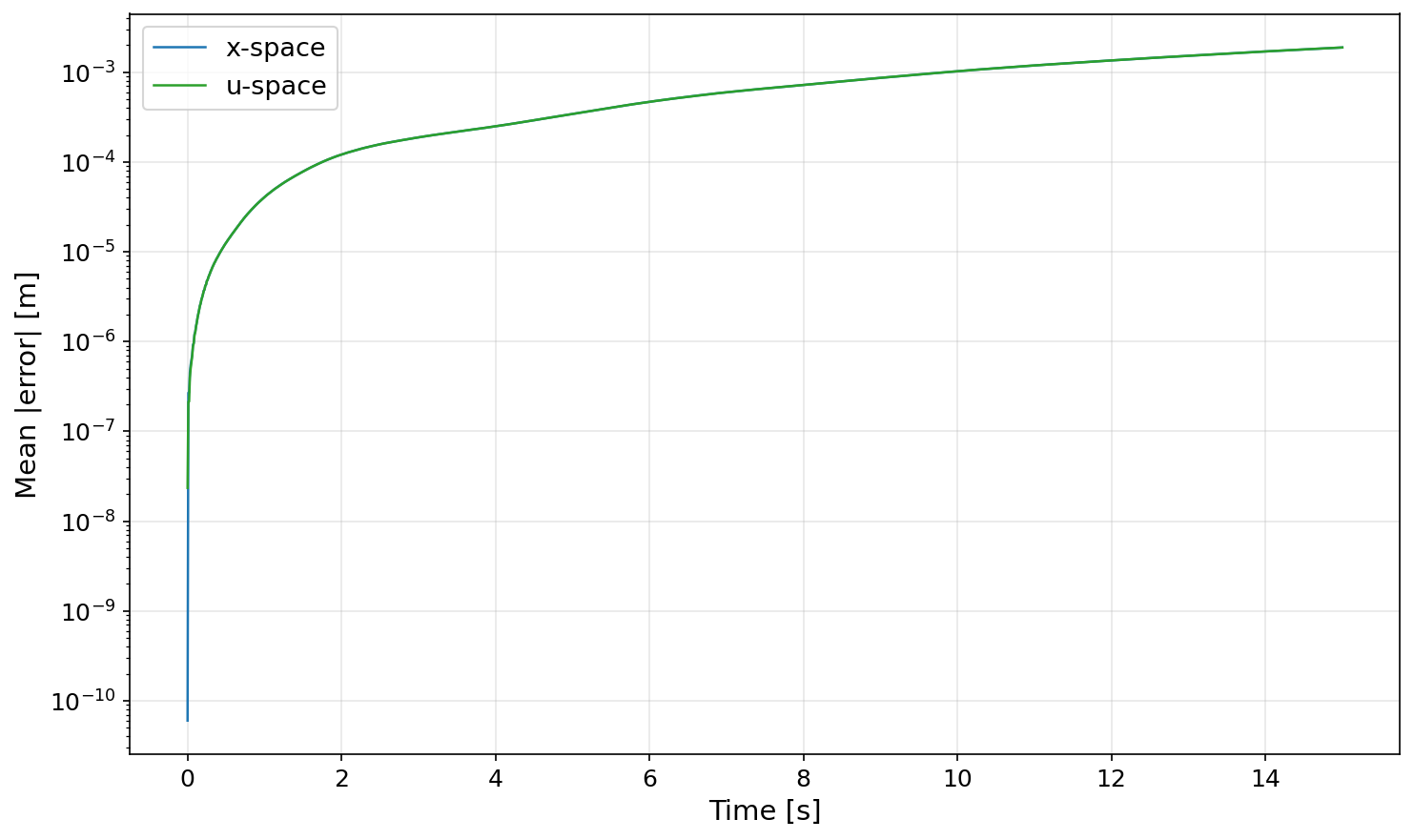}
        \caption{AB4}
        \label{fig:2nd_growth_consumer_ab4}
    \end{subfigure}
    \caption{Second-order simulation error growth for the navigation grade sensor for AB1 (a) and AB4 (b) methods. Both approaches show a reduced u-space error in the initial steps when compared to x-space, yet with the quadratic growth error growth dominating the higher-order method.}
    \label{fig:2nd_growth_consumer}
\end{figure}
\subsection{Simulation Results}
\label{sec:sim_results}
For each sensor type simulated measurements were generated and forcing functions were constructed for both first-order quaternion system using (\ref{eq:ak_forcing_def}) and for second order position-velocity system using (\ref{eq:pk_forcing}). For each system the solution was first propagated using x-space and then using u-space, and both results were compared against the simulated GT.
\subsubsection{\textbf{First-order System}}
\label{sec:sim_1st}
Table~\ref{tab:sim_1st_results} presents selected results for the strategic grade sensor. Even though the closed form supports all AB orders, the results show that the u-space performance saturates at AB1. This is due to two compounding mechanisms. First, unlike the second-order case where $\Phi$ depends on $\mathbf{f}_{q}(t)$ directly, the first-order closed form depends on $\mathbf{\dot{f}}_{q}(t)$, which is estimated via numerical differentiation. This introduces a new error source that the LTE optimization cannot eliminate. Second, the first-order system exhibits linear rather than quadratic error growth, leading to a smaller truncation budget for higher AB orders to reduce. The two factors compound to limit the practical benefit of the proposed approach to AB1. However, despite being limited to AB1, the u-space formulation achieves improved accuracy across all x-space AB orders, with $M_{\text{eff}}$ reaching the maximum evaluated window for x-space AB1. In line with the theory, the comparison to x-space AB1 exhibits the largest gain, reaching 97\% at $M=100$ (1 [s]) and maintaining above 95\% at $M=1500$ (15 [s]). Higher-order methods show more moderate improvements, with AB4 ranging from 5\% to 14\% across the evaluated windows, converging by the end to the noise-floor. Fig.~\ref{fig:1st_growth_consumer} compares the error growth over time for the navigation grade sensor for x-space AB1 and AB4 compared to u-space AB1. The absolute improvement is pronounced under x-space AB1, where the lower-order method produces larger truncation error that the gauge can reduce, while under x-space AB4 the errors are the same, u-space achieves these results with an integrator 3 orders lower.
\subsubsection{\textbf{Second-order System}}
\label{sec:sim_2nd}
Table~\ref{tab:sim_2nd_results} presents selected results for the strategic grade sensor. Similar to the first-order case, the u-space formulation saturates to the noise floor at a lower order than x-space, achieving this with AB2 compared to x-space AB3. AB1 exhibits the largest gains, reaching 99\% at $M=1$ (0.01 [s]) and maintaining 96\% at $M=1500$ (15 [s]). Fig.~\ref{fig:2nd_growth_consumer} compares the error growth over time for the navigation grade sensor for AB1 and AB4. Compared to the first-order case the absolute improvements are larger but the quadratic error growth is more pronounced in the overall trend.
\subsection{Experiment Results}
\label{sec:exp_results}
The data was processed in three phases as described in Section \ref{sec:real_setup} and the propagated results were compared to the recorded GT to compute the performance metrics.

\subsubsection{\textbf{Orientation}}
\label{sec:exp_orientation}
\begin{figure}[!tb]
    \centering
    \begin{subfigure}[b]{\columnwidth}
        \includegraphics[width=\textwidth]{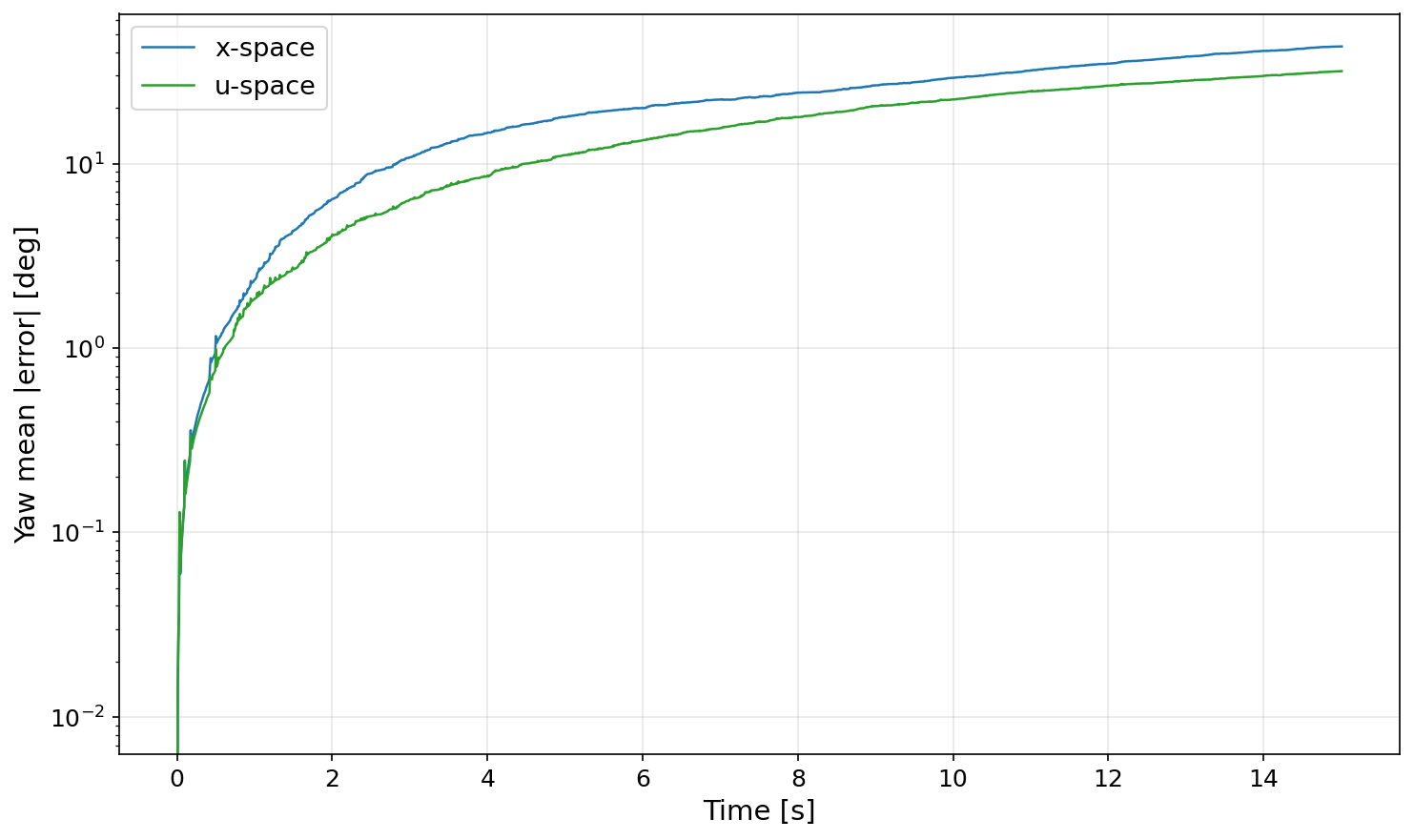}
        \caption{AB1}
        \label{fig:orientation_yaw_growth_ab1}
    \end{subfigure}
    \\[0.5em]
    \begin{subfigure}[b]{\columnwidth}
        \includegraphics[width=\textwidth]{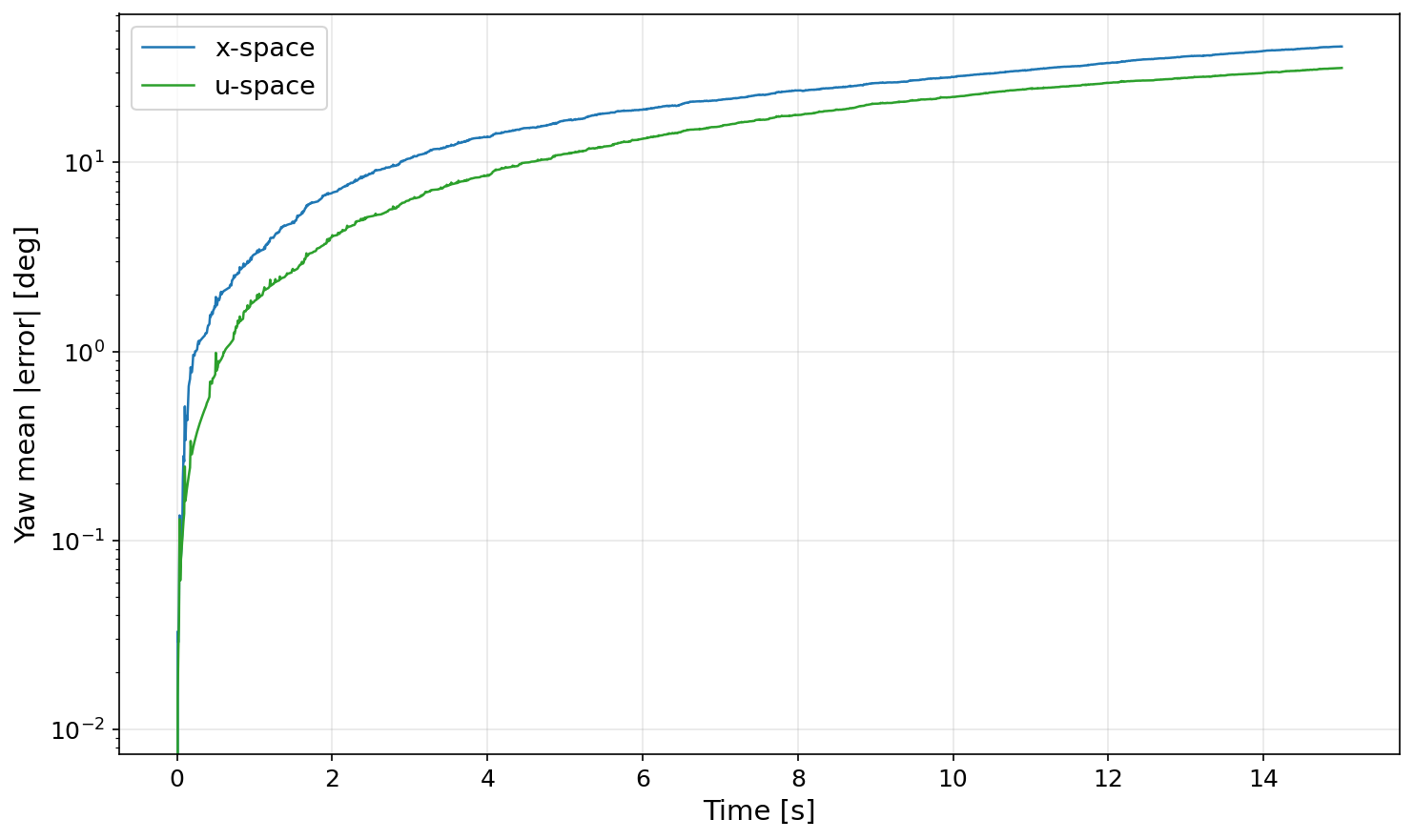}
        \caption{AB3}
        \label{fig:orientation_yaw_growth_ab3}
    \end{subfigure}
    \caption{Orientation experiment heading error growth for x-space AB1 (a) and AB3 (b) methods compared to u-space AB1. AB1 (u) maintains a consistent improvement across the entire trajectory.}
    \label{fig:orientation_yaw_growth}
\end{figure}

\begin{table*}[!tb]
    \centering
    \caption{Orientation experiment - geodesic error (RMSE) [$^\circ$], improvement [\%] relative to AB1 (x).}
    \label{tab:orientation}
    \begin{tabular}{l*{6}{c}c}
        \toprule
        & \multicolumn{6}{c}{\textbf{Prediction window} $M$ \textbf{[steps]}} & \\
        \cmidrule(lr){2-7}
        & 6 & 12 & 24 & 120 & 600 & 1800 & $\mathbf{M_{\text{eff}}}$ \\
        \midrule
        AB1 $x$-space & 3.7e-1 & 2.6e0 & 6.4e0 & 1.4e1 & 4.0e1 & 7.2e1 & $>$1800 ($>$15 [s]) \\
        AB2 $x$-space & 3.2e-1 & 6.7e-1 & 1.2e0 & 8.7e0 & 3.9e1 & 7.1e1 & $>$1800 ($>$15 [s]) \\
        AB3 $x$-space & 3.1e-1 & 6.5e-1 & 1.1e0 & 8.4e0 & 3.6e1 & 6.8e1 & $>$1800 ($>$15 [s]) \\
        AB4 $x$-space & 2.8e-1 & 5.9e-1 & 1.0e0 & 7.1e0 & 3.0e1 & 5.9e1 & --- \\
        \midrule
        AB1 $u$-space & \textbf{2.8e-1 (24\%)} & \textbf{5.9e-1 (77\%)} & \textbf{1.0e0 (84\%)} & \textbf{7.1e0 (49\%)} & \textbf{3.0e1 (25\%)} & \textbf{5.9e1 (18\%)} &  \\
        \bottomrule
    \end{tabular}
\end{table*}
Table~\ref{tab:orientation} presents the geodesic angle RMSE for the orientation experiment. Apart from x-space AB4, which converges to the noise floor as well, u-space outperforms x-space for the entire duration of the window, maintaining 13\% improvement at 15 [s] for AB3. Fig.~\ref{fig:orientation_yaw_growth} shows the heading error growth over time for AB1 and AB3, where despite the increased dynamics in this axis, the u-space curve remains below x-space for both cases.
\subsubsection{\textbf{Position with GT Quaternions}}
\label{sec:exp_pos_gt}
\begin{table*}[!tb]
    \centering
    \caption{Position experiment (GT quaternions) - 3D position error (RMSE) as a function of prediction window length [$m$].}
    \label{tab:position_gt_quat}
    \begin{tabular}{ll*{6}{c}c}
        \toprule
        & & \multicolumn{6}{c}{\textbf{Prediction window} $M$ \textbf{[steps]}} & \\
        \cmidrule(lr){3-8}
        & & 3 & 6 & 12 & 24 & 120 & 600 & $\mathbf{M_{\text{eff}}}$ \\
        \midrule
        \multirow{2}{*}{\textbf{AB1}}
        & RMSE$_x$ & 1.0e-4 & 2.7e-4 & 7.9e-4 & 2.7e-3 & 5.8e-2 & 1.3e0 & \\
        & RMSE$_u$ & \textbf{5.0e-5 (50\%)} & \textbf{1.7e-4 (37\%)} & \textbf{6.3e-4 (20\%)} & \textbf{2.4e-3 (11\%)} & 5.7e-2 (2\%) & 1.3e0 (0\%) & 24 (0.2 [s]) \\
        \midrule
        \multirow{2}{*}{\textbf{AB2}}
        & RMSE$_x$ & 7.4e-5 & 2.1e-4 & 6.9e-4 & 2.5e-3 & 5.8e-2 & 1.3e0 & \\
        & RMSE$_u$ & \textbf{4.9e-5 (34\%)} & \textbf{1.7e-4 (19\%)} & \textbf{6.3e-4 (9\%)} & \textbf{2.3e-3 (5\%)} & 5.7e-2 (2\%) & 1.3e0 (0\%) & 24 (0.2 [s]) \\
        \midrule
        \multirow{2}{*}{\textbf{AB3}}
        & RMSE$_x$ & 6.9e-5 & 1.9e-4 & 6.5e-4 & 2.5e-3 & 5.7e-2 & 1.3e0 & \\
        & RMSE$_u$ & \textbf{4.2e-5 (39\%)} & \textbf{1.7e-4 (11\%)} & \textbf{6.1e-4 (6\%)} & \textbf{2.3e-3 (5\%)} & 5.7e-2 (0\%) & 1.3e0 (0\%) & 24 (0.2 [s]) \\
        \midrule
        \multirow{2}{*}{\textbf{AB4}}
        & RMSE$_x$ & 6.5e-5 & 1.7e-4 & 6.2e-4 & 2.5e-3 & 5.7e-2 & 1.3e0 & \\
        & RMSE$_u$ & \textbf{3.7e-5 (43\%)} & \textbf{1.5e-4 (12\%)} & \textbf{5.8e-4 (6\%)} & \textbf{2.3e-3 (5\%)} & 5.7e-2 (0\%) & 1.3e0 (0\%) & 24 (0.2 [s]) \\
        \bottomrule
    \end{tabular}
\end{table*}
To isolate the gauge effect on the position integration, the body-to-navigation rotation was performed using GT quaternions. Table~\ref{tab:position_gt_quat} presents the 3D position RMSE. The results follow the general trend established in the simulation yet with a substantially reduced effective range due to the increased noise floor. AB1 achieves the largest improvement (50\% at $M=1$) with $M_{\text{eff}}=24$ (0.2 seconds) with higher-order methods experiencing reduced benefits but the same capped effective range.
\subsubsection{\textbf{Position with Propagated Quaternions}}
\label{sec:exp_pos_prop}
\begin{figure}[!tb]
    \centering
    \begin{subfigure}[b]{\columnwidth}
        \includegraphics[width=\textwidth]{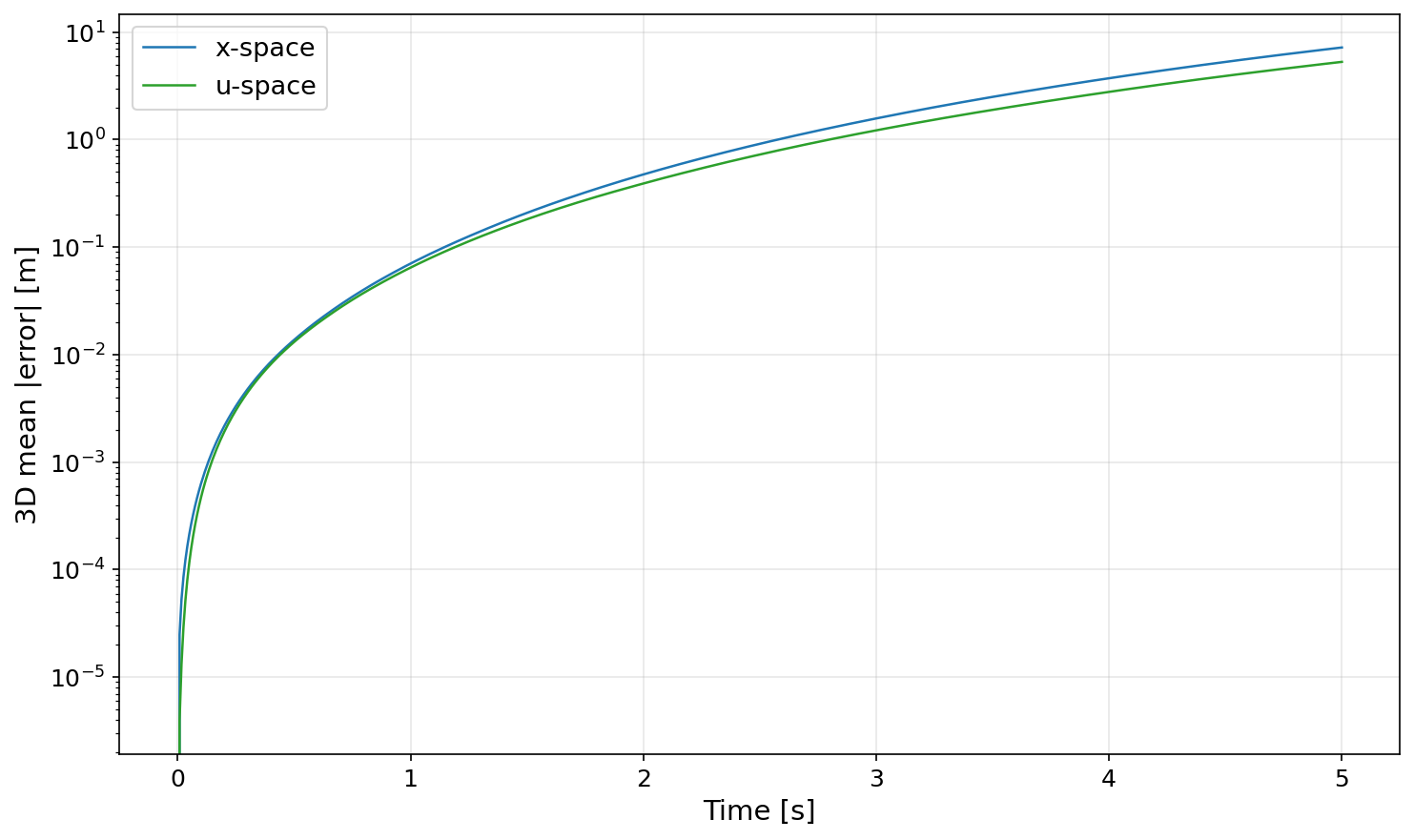}
        \caption{AB1}
        \label{fig:position_prop_growth_ab1}
    \end{subfigure}
    \\[0.5em]
    \begin{subfigure}[b]{\columnwidth}
        \includegraphics[width=\textwidth]{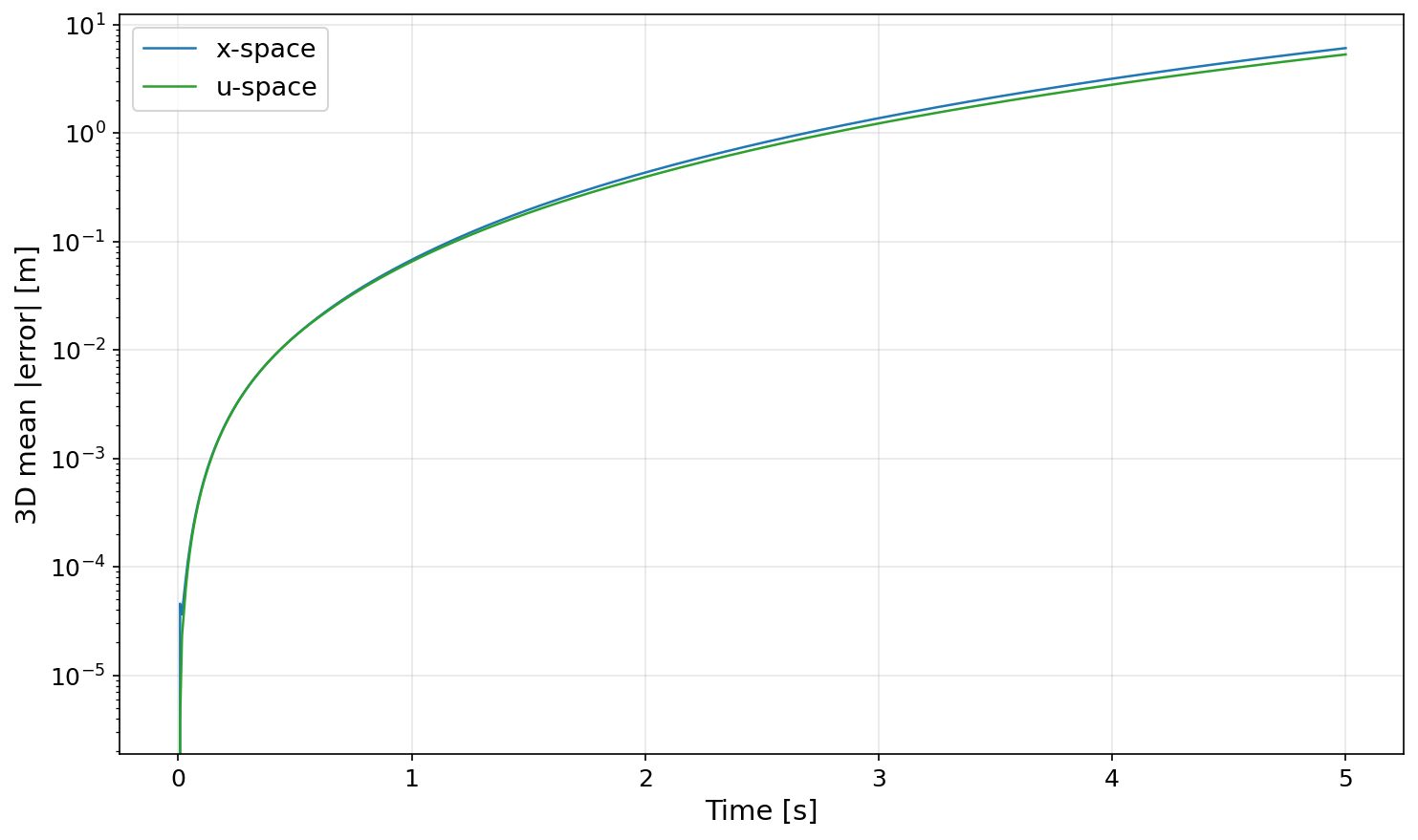}
        \caption{AB3}
        \label{fig:position_prop_growth_ab3}
    \end{subfigure}
    \caption{Position experiment (propagated quaternions) 3D error growth for AB1 (a) and AB3 (b). Both AB1 and AB3 maintain a visible separation over the duration as the orientation improvement compounds into position.}
    \label{fig:position_prop_growth}
\end{figure}

\begin{table*}[!tb]
    \centering
    \caption{Position experiment (propagated quaternions) - 3D position RMSE [$m$].}
    \label{tab:position_prop_quat}
    \begin{tabular}{ll*{6}{c}c}
        \toprule
        & & \multicolumn{6}{c}{\textbf{Prediction window} $M$ \textbf{[steps]}} & \\
        \cmidrule(lr){3-8}
        & & 3 & 6 & 12 & 24 & 120 & 600 & $\mathbf{M_{\text{eff}}}$ \\
        \midrule
        \multirow{2}{*}{\textbf{AB1}}
        & RMSE$_x$ & 1.0e-4 & 2.7e-4 & 8.2e-4 & 4.4e-3 & 2.7e-1 & 1.6e1 & \\
        & RMSE$_u$ & \textbf{4.9e-5 (51\%)} & \textbf{1.9e-4 (30\%)} & \textbf{6.8e-4 (17\%)} & \textbf{2.8e-3 (36\%)} & \textbf{1.2e-1 (56\%)} & \textbf{1.1e1 (31\%)} & $>$1800 ($>$15 [s]) \\
        \midrule
        \multirow{2}{*}{\textbf{AB2}}
        & RMSE$_x$ & 7.4e-5 & 2.2e-4 & 7.8e-4 & 3.0e-3 & 1.3e-1 & 1.3e1 & \\
        & RMSE$_u$ & \textbf{4.7e-5 (36\%)} & \textbf{1.9e-4 (14\%)} & \textbf{6.8e-4 (13\%)} & \textbf{2.8e-3 (7\%)} & \textbf{1.2e-1 (8\%)} & \textbf{1.1e1 (15\%)} & 1200 (10 [s]) \\
        \midrule
        \multirow{2}{*}{\textbf{AB3}}
        & RMSE$_x$ & 7.0e-5 & 2.0e-4 & 7.5e-4 & 2.9e-3 & 1.3e-1 & 1.2e1 & \\
        & RMSE$_u$ & \textbf{4.0e-5 (43\%)} & \textbf{1.8e-4 (10\%)} & \textbf{6.8e-4 (9\%)} & \textbf{2.7e-3 (5\%)} & \textbf{1.2e-1 (8\%)} & \textbf{1.1e1 (8\%)} & 600 (5 [s]) \\
        \midrule
        \multirow{2}{*}{\textbf{AB4}}
        & RMSE$_x$ & 6.6e-5 & 1.8e-4 & 7.0e-4 & 2.7e-3 & 1.2e-1 & 1.1e1 & \\
        & RMSE$_u$ & \textbf{3.7e-5 (44\%)} & \textbf{1.5e-4 (17\%)} & \textbf{6.3e-4 (10\%)} & \textbf{2.5e-3 (5\%)} & 1.2e-1 (0\%) & 1.1e1 (0\%) & 24 (0.2 [s]) \\
        \bottomrule
    \end{tabular}
\end{table*}
In this phase, the GT quaternions were replaced by the propagated quaternions from Section~\ref{sec:exp_orientation}, yielding a full inertial propagation pipeline for both state-spaces. Table~\ref{tab:position_prop_quat} presents the 3D position RMSE and Fig.~\ref{fig:position_prop_growth} shows the error growth for AB1 and AB4. While the total error has increased due to acceleration leakage from the quaternion misalignment error, this led to an increased relative improvement and an increased effective range, especially for AB1, which saw its effective range fully restored. AB2-4 saw a lower yet significant $\times$25-$\times$50 increase in duration due to the reduced error in the quaternion states.
\subsubsection{\textbf{Computational Footprint}}
The u-space formulation augments the state vector by a single state per channel relative to x-space, increasing the position system from two states ($p$, $v$) to three ($u_{p,1}$, $u_{p,2}$, $\Phi_p$) and the quaternion system from one state ($q$) to two ($u_{q,1}$, $u_{q,2}$). For an order-$k$ Adams-Bashforth integrator, each propagated state requires a cache of its $k$ most recent evaluations, so both the memory footprint and the per-step computational cost scale as $\mathcal{O}(S\,k)$, where $S$ is the number of states. At a fixed integrator order, the added state therefore raises the cost by a factor of $\times 1.5$ for the position-velocity system and a factor of $\times 2$ for the orientation system.

This fixed-order comparison, however, understates the practical picture. Since u-space attains a given accuracy at a lower order than x-space, for practical implementation we can perform an accuracy-matched comparison. Under this comparison, the relevant cost is that of the lowest order reaching the target error. For the orientation system, AB1 u-space matches AB4 x-space, leading to a 50\% footprint reduction. Likewise, for the position system, AB2 u-space matches AB3 x-space, giving the same cached evaluations in both cases. Meaning that for practical implementation, u-space is more computationally efficient than x-space, reaching a matched accuracy with a 24\% footprint reduction in the full inertial pipeline.
\subsection{Summary} 
\begin{figure}[!tb]
    \centering
    % fig_error_mechanism.tex
% Usage in main.tex:
%   \begin{figure}[t]
%       \centering
%       \input{files/flowcharts/fig_error_mechanism}
%       \caption{Error growth mechanism. The local truncation error
%       in the quaternion integration propagates outward through the
%       state-dependent forcing function and body-to-navigation
%       transformation, cascading into acceleration misalignment and
%       elevated position error.}
%       \label{fig:error_mechanism}
%   \end{figure}

\begin{tikzpicture}[
    font=\footnotesize\sffamily
]

% ── Concentric rings (outer to inner) ───────────────────────────
% Outer ring: position error
\fill[green!10] (0,0) ellipse (3.6cm and 3.0cm);
\draw[black!70, line width=1.2pt] (0,0) ellipse (3.6cm and 3.0cm);

% Middle ring: acceleration misalignment
\fill[orange!12] (0,0) ellipse (2.7cm and 2.2cm);
\draw[black!70, line width=1.0pt] (0,0) ellipse (2.7cm and 2.2cm);

% Inner-middle ring: model / forcing function error
\fill[red!10] (0,0) ellipse (1.8cm and 1.45cm);
\draw[black!70, line width=0.8pt] (0,0) ellipse (1.8cm and 1.45cm);

% Core: local truncation error
\fill[blue!15] (0,0) ellipse (0.95cm and 0.75cm);
\draw[black!70, line width=0.6pt] (0,0) ellipse (0.95cm and 0.75cm);

% ── Labels ───────────────────────────────────────────────────────
% Core label
\node[align=center, font=\scriptsize\sffamily] at (0, 0)
    {Truncation\\[-1pt]error};

% Inner-middle label
\node[align=center, font=\scriptsize\sffamily] at (0, -1.15)
    {Quaternion\\[-1pt]state error};

% Middle label
\node[align=center, font=\scriptsize\sffamily] at (0, -1.85)
    {Acceleration\\[-1pt]misalignment};

% Outer label
\node[align=center, font=\scriptsize\sffamily] at (0, -2.6)
    {Position error};

% ── Upward arrows between rings ─────────────────────────────────
% Arrow from core to inner-middle
\draw[->, >=Stealth, line width=0.7pt, black!90]
    (0, 0.78) -- (0, 1.15)
    node[right, font=\tiny\sffamily, xshift=1pt] {$\delta\mathbf{q}$};

% Arrow from inner-middle to middle
\draw[->, >=Stealth, line width=0.7pt, black!90]
    (0, 1.48) -- (0, 1.88)
    node[right, font=\tiny\sffamily, xshift=1pt] {$\delta\mathbf{f}$};

% Arrow from middle to outer
\draw[->, >=Stealth, line width=0.7pt, black!90]
    (0, 2.23) -- (0, 2.68)
    node[right, font=\tiny\sffamily, xshift=1pt] {$\delta\mathbf{p}$};

% ── Order-of-magnitude annotations ──────────────────────────────
% Core annotation (inside blue core, above label)
\node[font=\scriptsize\sffamily\bfseries] at (0, 0.42)
    {$\boldsymbol{\sim\!0.1}$ [m]};

% Outer annotation (above outer ring, near top arrow)
\node[font=\scriptsize\sffamily\bfseries] at (0, 3.45)
    {$\boldsymbol{\sim\!5.0}$ [m]};

\end{tikzpicture}
    \caption{Error growth mechanism illustration for AB1 at $M=1800$ (15 [s]). A global truncation error analysis results in an estimated 0.1 [m] position error. However, the quaternion integration cascades through the state-dependent forcing function and the body-to-navigation transformation, resulting in 5 [m] of accumulated position error in practice.}
    \label{fig:error_mechanism}
\end{figure}
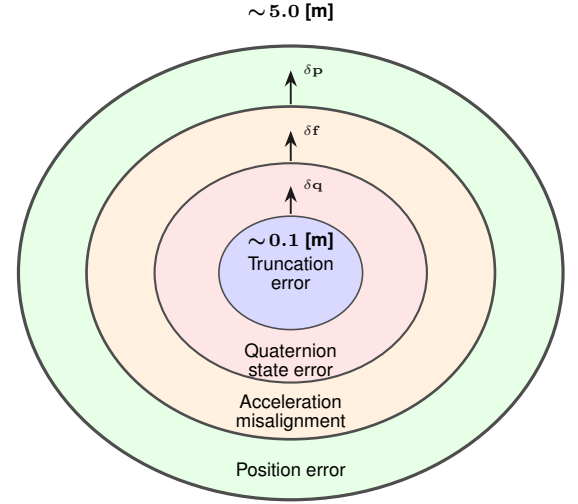

Sections~\ref{sec:exp_orientation} and \ref{sec:exp_pos_prop} showed increased performance when compared to their simulation counterparts, despite the increased noise level in the recorded IMU. This demonstrates a discrepancy between the theoretical and simulated truncation error analysis and its real-world effect. The theoretical analysis considers the truncation error in isolation. The local truncation error estimation assumes perfect initial conditions and examines the error growth in a single step, while the global truncation error estimation considers the aggregated effect of multiple steps. However, similarly to other error sources, the major effect of the truncation error stems from its downstream propagation rather than its self-accumulation. Fig.~\ref{fig:error_mechanism} illustrates this mechanism. The truncation error in the quaternion equation introduces a quaternion state error. Since the quaternion forcing function is state-dependent, this in turn leads to a model error between the approximated and the real forcing functions, which feeds through the body-to-navigation transformation into the acceleration misalignment error, resulting in an elevated position error. For low-order methods such as AB1, the downstream accumulation accounts for approximately a third of the total error budget and its reduction is enabled by the u-space.
\section{Conclusions}
\label{sec:conclusions}
In this paper we introduced the u-space formulation, a novel state-space formulation utilizing gauge freedom to reduce the numerical truncation error without modifying the integrator order or step size. The optimal gauge was derived in closed form for second-order position and velocity state propagation and adapted to first-order quaternion propagation using a data-driven method.
The proposed approach was evaluated through Monte Carlo simulations across four forcing functions, five sensor grades, and four Adams-Bashforth orders, as well as through a real-world inertial navigation experiment comprising 17 vehicle trajectories. The results showed consistent improvement across all tested conditions, with the largest gains observed where truncation error dominated the error budget (short prediction windows, higher sensor grades, and lower-order methods). For first-order systems, u-space AB1 achieved accuracy comparable to x-space AB4, while for second-order systems u-space AB2 saturated to the noise floor at the same level as x-space AB3. In the real-world experiments the results were in line with the simulation predictions and in some cases improved upon them. The largest real-world gains were observed in the full inertial mechanization pipeline where truncation error cascades through state-dependent feedback.
As demonstrated in the error growth mechanism analysis, the truncation error in the quaternion integration propagates through the forcing function and the body-to-navigation transformation, producing position errors that exceed the theoretical estimation by an order of magnitude. The u-space formulation reduces this divergence, resulting in improved accuracy. For AB1, a 31\% position RMSE reduction was observed after 600 steps (5 seconds), with AB3 maintaining an 8\% reduction over the same duration.
The proposed approach is applicable to high-end inertial systems, where the sensor noise floor is low and truncation error constitutes a larger share of the total error budget, and to consumer-grade systems operating with high-frequency aiding updates, which at rates of approximately 5~Hz fall within the proposed approach's effective range. Examples for the former include marine navigation, airborne inertial navigation, and spacecraft navigation, while examples for the latter include GNSS-aided automotive navigation, pedestrian dead reckoning, and visual-inertial odometry on automotive platforms.
%
%\ifCLASSOPTIONcaptionsoff
%  \newpage
%\fi
\bibliographystyle{IEEEtran}
\bibliography{export}

@article{gurfil2006mitigating,
  title={{M}itigating the {I}ntegration {E}rror in {N}umerical {S}imulations of {N}ewtonian {S}ystems},
  author={Gurfil, Pini and Klein, Itzik},
  journal={International Journal for Numerical Methods in Engineering},
  volume={68},
  number={2},
  pages={267--297},
  year={2006},
  publisher={Wiley Online Library}
}

@article{yampolsky2024multiple,
  title={{M}ultiple and {G}yro-free {I}nertial {D}atasets},
  author={Yampolsky, Zeev and Stolero, Yair and Pri-Hadash, Nitsan and Solodar, Dan and Massas, Shira and Savin, Itai and Klein, Itzik},
  journal={Scientific Data},
  volume={11},
  number={1},
  pages={1080},
  year={2024},
  publisher={Nature Publishing Group UK London}
}

@article{papez2013numerical,
  title={{N}umerical {A}spects of {I}nertial {N}avigation},
  author={Papez, Milan and Pivonka, Petr},
  journal={IFAC Proceedings Volumes},
  volume={46},
  number={28},
  pages={262--267},
  year={2013},
  publisher={Elsevier}
}

@article{featherstone1983calculation,
  title={{T}he {C}alculation of {R}obot {D}ynamics {U}sing {A}rticulated-body {I}nertias},
  author={Featherstone, Roy},
  journal={The International Journal of Robotics Research},
  volume={2},
  number={1},
  pages={13--30},
  year={1983},
  publisher={Sage Publications Sage CA: Thousand Oaks, CA}
}

@article{nakai2024ordinary,
  title={{O}rdinary {D}ifferential {E}quation-based {MIMO} {S}ignal {D}etection},
  author={Nakai-Kasai, Ayano and Wadayama, Tadashi},
  journal={IEEE Transactions on Signal Processing},
  volume={72},
  pages={4147--4162},
  year={2024},
  publisher={IEEE}
}

@article{verlet1967computer,
  title={{C}omputer "{E}xperiments" on {C}lassical {F}luids. {I}. {T}hermodynamical {P}roperties of {L}ennard-{J}ones {M}olecules},
  author={Verlet, Loup},
  journal={Physical Review},
  volume={159},
  number={1},
  pages={98},
  year={1967},
  publisher={APS}
}

@article{dahlquist1956convergence,
  title={{C}onvergence and {S}tability in the {N}umerical {I}ntegration of {O}rdinary {D}ifferential {E}quations},
  author={Dahlquist, Germund},
  journal={Mathematica Scandinavica},
  pages={33--53},
  year={1956},
  publisher={JSTOR}
}

@article{dormand1980family,
  title={{A} {F}amily of {E}mbedded {R}unge-{K}utta {F}ormulae},
  author={Dormand, John R and Prince, Peter J},
  journal={Journal of Computational and Applied Mathematics},
  volume={6},
  number={1},
  pages={19--26},
  year={1980},
  publisher={Elsevier}
}

@article{richardson1911ix,
  title={{T}he {A}pproximate {A}rithmetical {S}olution by {F}inite {D}ifferences of {P}hysical {P}roblems {I}nvolving {D}ifferential {E}quations, with an {A}pplication to the {S}tresses in a {M}asonry {D}am},
  author={Richardson, Lewis Fry},
  journal={Philosophical Transactions of the Royal Society of London. Series A, Containing Papers of a Mathematical or Physical Character},
  volume={210},
  number={459-470},
  pages={307--357},
  year={1911},
  publisher={The Royal Society London}
}

@article{romberg1955vereinfachte,
  title={{S}implified {N}umerical {I}ntegration},
  author={Romberg, Werner},
  journal={Norske Vid. Selsk. Forh.},
  volume={28},
  pages={30--36},
  year={1955},
  publisher={Trondheim}
}

@article{butusov2021adaptive,
  title={{A}daptive {S}tepsize {C}ontrol for {E}xtrapolation {S}emi-implicit {M}ultistep {ODE} {S}olvers},
  author={Butusov, Denis},
  journal={Mathematics},
  volume={9},
  number={9},
  pages={950},
  year={2021},
  publisher={MDPI}
}

@article{sanz1992symplectic,
  title={{S}ymplectic {I}ntegrators for {H}amiltonian {P}roblems: an {O}verview},
  author={Sanz-Serna, Jesus M},
  journal={Acta Numerica},
  volume={1},
  pages={243--286},
  year={1992},
  publisher={Cambridge University Press}
}

@article{hochbruck2010exponential,
  title={{E}xponential {I}ntegrators},
  author={Hochbruck, Marlis and Ostermann, Alexander},
  journal={Acta Numerica},
  volume={19},
  pages={209--286},
  year={2010},
  publisher={Cambridge University Press}
}

@article{blanes2009magnus,
  title={{T}he {M}agnus {E}xpansion and {S}ome of {I}ts {A}pplications},
  author={Blanes, Sergio and Casas, Fernando and Oteo, Jose-Angel and Ros, Jos{\'e}},
  journal={Physics Reports},
  volume={470},
  number={5-6},
  pages={151--238},
  year={2009},
  publisher={Elsevier}
}

@article{gurfil2007stabilizing,
  title={{S}tabilizing the {E}xplicit {E}uler {I}ntegration of {S}tiff and {U}ndamped {L}inear {S}ystems},
  author={Gurfil, Pini and Klein, Itzik},
  journal={Journal of Guidance, Control, and Dynamics},
  volume={30},
  number={6},
  pages={1659--1667},
  year={2007}
}

@inproceedings{alghamdi2025autonomous,
  title={{A}utonomous {N}avigation {S}ystems in {GPS}-denied {E}nvironments: {A} {R}eview of {T}echniques and {A}pplications},
  author={Alghamdi, Saleh and Alahmari, Sultan and Yonbawi, Saud and Alsaleem, Khalid and Ateeq, Fahad and Almushir, Faris},
  booktitle={2025 11th International Conference on Automation, Robotics, and Applications (ICARA)},
  pages={290--299},
  year={2025},
  organization={IEEE}
}

@article{giroux2003validation,
  title={{V}alidation and {P}erformance {E}valuation of a {S}imulink {I}nertial {N}avigation {S}ystem {S}imulator},
  author={Giroux, Richard and Leach, BW and Landry Jr, R and Gourdeau, Richard},
  journal={Canadian Aeronautics and Space Journal},
  volume={49},
  number={4},
  pages={149--161},
  year={2003},
  publisher={Canadian Aeronautics and Space Institute}
}

@book{titterton2004strapdown,
  title={{S}trapdown {I}nertial {N}avigation {T}echnology},
  author={Titterton, DH},
  publisher={The Institution of Engineering and Technology},
  year={2004}
}

@book{groves2008,
   title = {{P}rinciples of {G}{N}{S}{S} {I}nertial and {M}ulti-{S}ensor {I}ntegrated {N}avigation {S}ystems - {G}{N}{S}{S} {T}echnology and {A}pplications},
   author = {Groves, Paul D.},
   year = {2015},
   publisher = {Norwood, MA, USA: Artech House},
}

@book{farrell2008aided,
  title={{A}ided {N}avigation: {G}{P}{S} with {H}igh {R}ate {S}ensors},
  author={Farrell, Jay},
  year={2008},
  publisher={McGraw-Hill, Inc.}
}

@book{boyce2021elementary,
  title={{E}lementary {D}ifferential {E}quations and {B}oundary {V}alue {P}roblems},
  author={Boyce, William E and DiPrima, Richard C and Meade, Douglas B},
  year={2021},
  publisher={John Wiley \& Sons}
}

@book{tenenbaum1985ordinary,
  title={{O}rdinary {D}ifferential {E}quations: an {E}lementary {T}extbook for {S}tudents of {M}athematics, {E}ngineering, and the {S}ciences},
  author={Tenenbaum, Morris and Pollard, Harry},
  year={1985},
  publisher={Courier Corporation}
}

@book{coddington1956theory,
  title={{T}heory of {O}rdinary {D}ifferential {E}quations},
  author={Coddington, Earl A and Levinson, Norman and Teichmann, T},
  year={1956},
  publisher={American Institute of Physics}
}

@book{iserles2009first,
  title={{A} {F}irst {C}ourse in the {N}umerical {A}nalysis of {D}ifferential {E}quations},
  author={Iserles, Arieh},
  number={44},
  year={2009},
  publisher={Cambridge University Press}
}

@book{ogata2010modern,
  title={{M}odern {C}ontrol {E}ngineering},
  author={Ogata, Katsuhiko},
  year={2010},
  publisher={Prentice hall}
}

@book{butcher2016numerical,
  title={{N}umerical {M}ethods for {O}rdinary {D}ifferential {E}quations},
  author={Butcher, John Charles},
  year={2016},
  publisher={John Wiley \& Sons}
}

@book{bashforth1883attempt,
  title={{A}n {A}ttempt to {T}est the {T}heories of {C}apillary {A}ction by {C}omparing the {T}heoretical and {M}easured {F}orms of {D}rops of {F}luid},
  author={Bashforth, Francis and Adams, John Couch},
  year={1883},
  publisher={University Press}
}

@article{zeinali2024imunet,
  title={{IMUN}et: {E}fficient {R}egression {A}rchitecture for {I}nertial {IMU} {N}avigation and {P}ositioning},
  author={Zeinali, Behnam and Zanddizari, Hadi and Chang, Morris J},
  journal={IEEE Transactions on Instrumentation and Measurement},
  volume={73},
  pages={1--13},
  year={2024},
  publisher={IEEE}
}

@article{hurwitz2024deep,
  title={{D}eep-learning-assisted {I}nertial {D}ead {R}eckoning and {F}usion},
  author={Hurwitz, Dror and Cohen, Nadav and Klein, Itzik},
  journal={IEEE Transactions on Instrumentation and Measurement},
  volume={74},
  pages={1--9},
  year={2024},
  publisher={IEEE}
}

@article{al2023imu,
  title={{IMU} {H}and {C}alibration for {L}ow-cost {MEMS} {I}nertial {S}ensors},
  author={Al Jlailaty, Hussein and Celik, Abdulkadir and Mansour, Mohammad M and Eltawil, Ahmed M},
  journal={IEEE Transactions on Instrumentation and Measurement},
  volume={72},
  pages={1--16},
  year={2023},
  publisher={IEEE}
}

@article{xu2024nonlinearity,
  title={{N}onlinearity-aware {ZUPT}-aided {P}edestrian {I}nertial {N}avigation {B}ased on {C}ubature {K}alman {F}ilter in {U}rban {C}anyons},
  author={Xu, Ruijie and Chen, Shichao and Bai, Shiyu and Wen, Weisong},
  journal={IEEE Transactions on Instrumentation and Measurement},
  volume={73},
  pages={1--15},
  year={2024},
  publisher={IEEE}
}

@misc{XDOT,
  title = {{Xsens DOT}},
  howpublished = "\url{https://www.movella.com/products/wearables/movella-dot}",
  note = "[Accessed Jun. 15, 2026]"
}

@misc{InertialLabs,
  title = {{Inertial Labs}},
  howpublished = "\url{https://inertiallabs.com/products/mru-ws-motion-reference-units-and-wave-sensors/}",
  note = "[Accessed Jun. 15, 2026]"
}
% biography section
\end{document}